\documentclass[prl,amsmath,twocolumn,amssymb,aps,superscriptaddress,floatfix,preprintnumbers,10pt,notitlepage]{revtex4-1}
\usepackage{times}
\usepackage{graphicx}
\usepackage{amsmath, braket, amsfonts}
\usepackage{amssymb}
\usepackage{natbib}
\usepackage{sidecap}
\usepackage{bm, color, ulem}
\usepackage{xcolor}
\usepackage{lipsum}
\DeclareUnicodeCharacter{03BD}{{$\nu$}}

\begin{document}

\title{Half and quarter metals in rhombohedral trilayer graphene}
\author{Haoxin Zhou}
\thanks{These authors contributed equally to this work}
\affiliation{Department of Physics, University of California at Santa Barbara, Santa Barbara CA 93106, USA}
\author{Tian Xie}
\thanks{These authors contributed equally to this work}
\affiliation{Department of Physics, University of California at Santa Barbara, Santa Barbara CA 93106, USA}
\author{Areg Ghazaryan}
\affiliation{IST Austria, Am Campus 1, 3400 Klosterneuburg, Austria}
\author{Tobias Holder}
\affiliation{Department of Condensed Matter Physics, Weizmann Institute of Science, Rehovot 76100, Israel}
\author{James R. Ehrets}
\affiliation{Department of Physics, University of California at Santa Barbara, Santa Barbara CA 93106, USA}
\email{Current address: Harvard}
\author{Eric M. Spanton}
\affiliation{Department of Physics, University of California at Santa Barbara, Santa Barbara CA 93106, USA}
\email{Current address: Michigan}
\author{Takashi Taniguchi}
\affiliation{International Center for Materials Nanoarchitectonics,
National Institute for Materials Science,  1-1 Namiki, Tsukuba 305-0044, Japan}
\author{Kenji Watanabe}
\affiliation{Research Center for Functional Materials,
National Institute for Materials Science, 1-1 Namiki, Tsukuba 305-0044, Japan}
\author{Erez Berg}
\affiliation{Department of Condensed Matter Physics, Weizmann Institute of Science, Rehovot 76100, Israel}
\author{Maksym Serbyn}
\affiliation{IST Austria, Am Campus 1, 3400 Klosterneuburg, Austria}
\author{Andrea F. Young}
\email{andrea@physics.ucsb.edu}
\affiliation{Department of Physics, University of California at Santa Barbara, Santa Barbara CA 93106, USA}
\date{\today}
\begin{abstract}
\end{abstract}
\maketitle

\textbf{Ferromagnetism is most common in transition metal compounds where electrons occupy highly localized d-orbitals.  
However, ferromagnetic order may also arise in low-density two-dimensional electron systems\cite{tanatar_ground_1989}, with signatures observed in silicon\cite{pudalov_thermodynamic_2015}, III-V semiconductor systems\cite{hossain_observation_2020-1}, and graphene moir\'e heterotructures\cite{sharpe_emergent_2019,chen_tunable_2020}.  
Here we show that gate-tuned van Hove singularities in rhombohedral trilayer graphene drive the spontaneous ferromagnetic polarization of the electron system into one or more spin- and valley flavors. 
Using capacitance measurements on graphite-gated van der Waals heterostructures, we find a cascade of density- and electronic displacement field tuned phase transitions marked by negative electronic compressibility.  
The transitions define the boundaries between phases in which quantum oscillations have either four-fold, two-fold, or one-fold degeneracy, associated with a spin and valley degenerate normal metal, spin-polarized `half-metal', and spin and valley polarized `quarter metal', respectively. 
For electron doping, the salient features of the data are well captured by a phenomenological Stoner model\cite{stoner_collective_1938} that includes a valley-anisotropic Hund's coupling, likely arising from electron-electron interactions at the scale of  the graphene lattice.
For hole filling, the single particle band structure features a finite-density van Hove singularity, and we observe a richer phase diagram featuring a delicate interplay of broken symmetries and transitions in the Fermi surface topology. 
Finally, we introduce a moir\'e superlattice by rotational alignment of a hexagonal boron nitride substrate\cite{chen_evidence_2019,chen_tunable_2020}. 
Remarkably, we find that the superlattice perturbs the preexisting isospin order only weakly, leaving the basic phase diagram intact while catalyzing the formation of topologically nontrivial gapped states whenever itinerant half- or quarter metal states occur at half- or quarter superlattice band filling.  Our results show that rhombohedral graphene is an ideal platform for well controlled tests of many-body theory, and reveal magnetism in moir\'e materials\cite{sharpe_emergent_2019,serlin_intrinsic_2020,chen_tunable_2020,polshyn_electrical_2020} to be fundamentally itinerant in nature.}   

\begin{figure}
\centering
\includegraphics[width=\columnwidth]{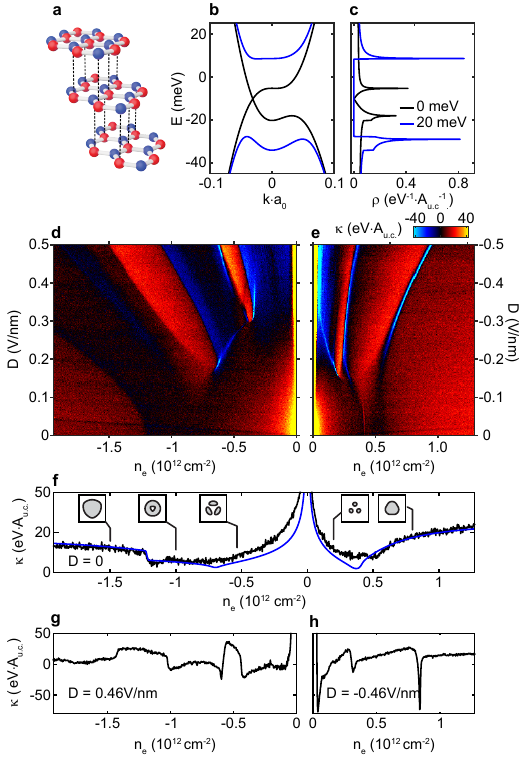}  
\caption{\textbf{Spontaneous symmetry breaking in rhombohedral trilayer graphene}. 
\textbf{a}, Crystal structure of rhombohedral trilayer graphene. 
\textbf{b}, Band structure of rhombohedral trilayer graphene with an interlayer potential $\Delta_1=0$ (black) and 30\,meV (blue) calculated from the six-band continuum model (see S.I.). Here $a_0$=.246nm is the graphene lattice constant. 
\textbf{c}, Corresponding single-particle density of states $\rho$ versus energy at zero temperature. 
\textbf{d}, False-color plot of the inverse compressibility as a function of displacement field and carrier density for hole doping and  
 \textbf{e}, electron doping.
\textbf{f}, Black: inverse compressibility measured at $D=0$. Blue: inverse compressibility calculated from the single particle, six-band continuum model. 
Insets: Fermi contours calculated from the continuum model (see also Fig. \ref{fig:S:FFT_at_p0=0}). Countours are plotted as a function of wave vector near one of the corners of the Brillouin zone.  The wave vector components $(k_x,k_y)$ range between $\pm 0.05/a_{\rm 0}$ in all plots.
\textbf{g}~Inverse compressibility as a function of carrier density measured at $D=0.46$\,V/nm and 
and \textbf{h},  $D=-0.46$\,V/nm.
}\label{fig1}
\end{figure}

\begin{figure*}
\centering
\includegraphics[width=160mm]{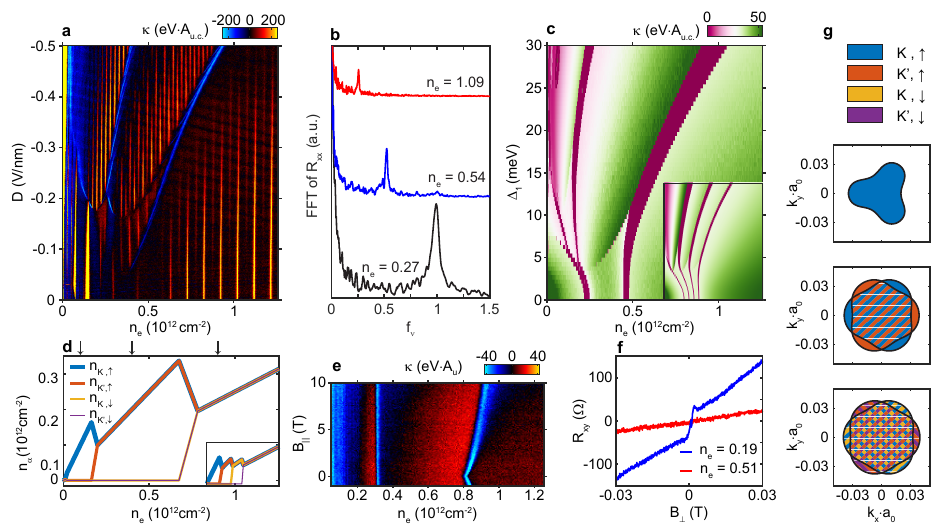}
\caption{\textbf{Stoner ferromagnetism in the conduction band.}
\textbf{a}, 
False-color plot of the inverse compressibility $\kappa$ measured at $B_{\perp}=1$\,T.
\textbf{b}, 
Fourier transform of $R_{xx}(1/B)$ 
at $D=-0.43$~V/nm and the indicated values of $n_{\rm e}$, expressed in units of $\times10^{12} {\rm cm^{-2}}$.
Data are plotted as a function of $f_\nu={f_{B}}/({\phi_0 n_e})$, where $f_{B}$ is the oscillation frequency (measured in Tesla) and $\phi_0=h/e$ is the magnetic flux quantum.  
$f_\nu$ can be interpreted as the fraction of the total Fermi surface area enclosed by a given orbit.  
\textbf{c}, Inverse compressibility at $B=0$ calculated from the Stoner model described in the main text with $U=30$\,eV, $J=-9$\,eV. Inset: Same results with $U=30$\,eV, $J=0$.
\textbf{d}, Partial densities of each spin-valley flavor density calculated from the Stoner model at a fixed interlayer potential of 20 meV for parameters in panel c. 
\textbf{e}, In-plane magnetic field dependence of the $\kappa$ measured at $D = -0.43$\,V/nm.
\textbf{f}, Hall resistivity $R_{xy}$ measured at $n_{\rm e} = 0.19\times10^{12}$~cm$^{-2}$
(blue) and $n_{\rm e}=0.51\times10^{12}$\,cm$^{-2}$ (red). Both curves are measured at $D=-0.4\,$V/nm and $T=0.5\,$K.
\textbf{g}, Calculated Fermi contours of each flavor corresponding to the three carrier density values indicated by the arrows in panel d. The Fermi contours are color coded to distinguish spin- and valley- flavors.
}\label{fig2}
\end{figure*}

In metals, electronic interaction effects become dominant when the potential energy arising from the Coulomb repulsion becomes larger than the kinetic energy the electrons inherit from the band dispersion. 
This condition is captured by the Stoner criterion\cite{stoner_collective_1938}, $UD_{\rm F}>1$, which links the strength of the Coulomb repulsion, $U$, and the single particle density of electronic states  at the Fermi energy, $D_{\rm F}$. 
The Stoner criterion provides the qualitative basis for ferromagnetism in transition metals, where a partially filled narrow band arising from the highly localized d-orbitals contributes a large density of states at the Fermi level. 
However, quantitative modeling of ferromagnetism in metals relies heavily on approximate treatments of exchange and correlation effects, which are typically difficult to directly benchmark to experiment.  
Model systems such as the two dimensional electron gas offer a path forward, providing a well controlled venue to benchmark many body theory using both more elaborate numerical methods\cite{tanatar_ground_1989}, and, ideally, precision experiments. 
In this regard, graphene-based van der Waals heterostructures present a new opportunity due to their exceptionally low disorder, well known single particle band structures, and a high degree of in-situ control using electric and magnetic fields. 

In this letter, we focus on rhombohedrally stacked graphene trilayers, characterized by ``ABC'' stacking order where A, B, and C refer to inequivalent relative placements of individual graphene layers within the multilayer crystal (Fig.~\ref{fig1}a). 
All rhombohedral graphite multilayers feature van Hove singularities at or near the band edge\cite{zhang_band_2010} where the density of states diverges, which are enhanced by a perpendicular electric displacement field (Fig.~\ref{fig1}b-c). Low density ferromagnetism was predicted theoretically for bilayers shortly after the experimental isolation of graphene\cite{castro_low-density_2008}, but signatures of correlation physics at finite density have been observed only recently in rhombohedral trilayers\cite{lee_gate_2019}, tetralayers\cite{kerelsky_moireless_2021} and multilayers\cite{shi_electronic_2020}.  
Moreover, experiments on rhombohedral trilayer graphene aligned to hexagonal boron nitride\cite{chen_evidence_2019,chen_tunable_2020,chen_signatures_2019} have shown the emergence of a flat electronic band hosting correlation driven insulators appearing at integer filling of the moir\'e superlattice unit cell. 
However, a microscopic unified picture of these disparate effects has remained elusive.

\subsection{Phase diagram of rhombohedral trilayer graphene}
To probe the ground state thermodynamic properties of rhombohedral trilayers, we measure the penetration field capacitance\cite{eisenstein_negative_1992}, which is directly proportional to the inverse electronic compressibility $\kappa=\partial\mu/\partial n_{\rm e}$ (here $\mu$ is the chemical potential and $n_{\rm e}$ is the charge carrier density). 
We use dual graphite gated devices\cite{zibrov_tunable_2017}, which minimize the charge disorder, enabling us to probe the intrinsic properties of these structures as a function of both carrier density $n_{\rm e}$ and the applied perpendicular electric displacement field $D$ (see Fig. \ref{fig:S:fab} and Methods).  Fig.~\ref{fig1}d-e shows the inverse compressibility measured at zero magnetic field and the base temperature of our dilution refrigerator (temperature dependent data are presented in Fig. \ref{fig:S:capacitance}).  At all values of $D$, the inverse compressibility has a maximum at charge neutrality, consistent with the Dirac nodes (for $D=0$) or displacement field induced band gap (for $|D|>0$) expected from the single particle band structure.  At $D=0$, density dependent $\kappa$ is marked by only two significant features at finite density in the form of a sharp step at $n_{\rm e}\approx -1.2\times 10^{12}\,{\rm cm}^{-2}$ and a kink at $n_{\rm e}\approx 0.5\times 10^{12}\,{\rm cm}^{-2}$.  
These features are qualitatively captured by the tight binding band structure of Fig.~\ref{fig1}b-c, and correspond to van Hove singularities arising from Lifshitz transitions in the Fermi sea topology, which is simply connected at high densities but is described by an annulus or multiple pockets at lower densities.

While some evidence of correlation driven physics is evident for $D=0$ at the charge neutrality point (Fig. \ref{fig:S:transport_cnp}) and near $n_e\approx 0.4\times 10^{12}\,$cm$^{-2}$ (Fig. \ref{fig:S:FFT_at_p0=0}), deviations from the single-particle picture are most evident at high $|D|$ where experimental data show many features not present in the tight binding model, in which large displacement field simplifies the electronic structure (see S.I.).   
The experimental $\kappa$ features multiple regions of near constant compressibility separated by boundaries where $\kappa$ is strongly negative.  Negative compressibility is generally associated with electronic correlations\cite{eisenstein_negative_1992}, and may arise at first order phase transitions characterized by phase separation. For electron doping, the high $|D|$ phase diagram appears to consist of three distinct phases at low, intermediate, and high density separated by first order phase transitions, while for hole doping ($n_{\rm e}<0$) the phase diagram is more complex. In both cases, however, negative compressibility features develop at finite $|D|$ and evolve towards higher $|n_{\rm e}|$ with increasing $|D|$. These features rapidly wash out as the temperature is raised, though associated features remain visible at 5K (Fig. \ref{fig:S:capacitance}). 

The nature of competing phases is immediately revealed by finite magnetic field measurements, shown for electron doping in Fig.~\ref{fig2}a for a magnetic field $B_\perp=1$T applied perpendicular to the sample plane. At this field, energy gaps between Landau levels are easily visible as peaks in the inverse compressibility, while the phase boundaries are only slightly altered relative to the zero magnetic field case.  As is evident in Fig.~\ref{fig2}a  the phase transitions observed at $B=0$ separate regions of contrasting Landau level degeneracy. In the high density phase, the Landau levels have the combined four-fold degeneracy of the spin and valley flavors native to graphene systems; similarly, at low $n_{\rm e}$ and low $D$, a 12-fold symmetry emerges due to additional degeneracy of local minima in the strongly trigonally warped Fermi surface\cite{varlet_anomalous_2014,zibrov_emergent_2018}. 
However, in the intermediate and low-density phases, respectively, the degeneracy is reduced to two-fold and one-fold.  
This trend is evident in low-magnetic field magnetoresistance oscillations in the three regimes, Fourier transforms of which are shown in Fig.~\ref{fig2}b (see also Figs. \ref{fig:S:sdh_n_side} and Fig. \ref{fig:S:oscillation}). The loss of degeneracy is consistent with a zero magnetic field phase diagram which contains two distinct phases that spontaneously break the combined spin- and valley isospin symmetry. In this picture, the intermediate density phase consists of \textit{two} degenerate Fermi surfaces at $B_\perp=0$, constituting a `half-metal' as compared to the normally four-fold degenerate graphene, while the low-density phase has a single Fermi surface and is thus a `quarter-metal.'

\subsection{Stoner ferromagnetism}
To better understand the mechanisms leading to the rich magnetic phase diagram observed, we study a four-component Stoner model\cite{zondiner_cascade_2020}. Within this model, the grand potential  per unit area is given by
\begin{align}
    \frac{\Phi}{A} = &\sum_{\alpha}E_0(\mu_\alpha)+\frac{U A_{\rm u.c.}}{2}\sum_{\alpha \neq \beta}n_\alpha n_\beta+\mu \sum_\alpha{n_\alpha}.\nonumber
\end{align}
Here $A$, $A_{\rm u.c.}$ are the area of the sample and unit cell respectively, $\alpha$ and $\beta$ index the four spin- and valley flavors, and $n_\alpha$ and $\mu_\alpha$ are the density and chemical potential for a given flavor $\alpha$. The first term, with $E_0(\mu_\alpha)=\int_0^\mu \epsilon\rho(\epsilon)d\epsilon$, where $\rho(\epsilon)$ is a density of states per area, accounts for the kinetic energy, and is minimized by occupying all flavors equally.  The second term accounts for the effect of exchange interactions, whose strength is parameterized by a constant energy $U$ and which we assume to be symmetric within the spin- and valley isospin space. The exchange energy is minimized when fewer flavors are occupied.

The inverse compressibility calculated within this model (Figure \ref{fig2}c, inset) captures several key features of the experimental data.  First, the model produces a cascade of symmetry broken phases in which the degeneracy of the Fermi surface is reduced (see Fig. \ref{fig2}d, inset, and S. I.).  
Moreover, the phase transitions separating these phases follow trajectories in the $n_{\rm e}-D$ plane very similar to those observed experimentally.  This dependence can be directly related to the evolution of the band-edge van Hove singularities, which cause the Stoner criterion for ferromagnetism to be satisfied at ever higher $|n|$ with increasing $|D|$ as more states accumulate near the band edge. 

However, the model deviates significantly from the experimental data, predicting a three-fold degenerate phase that is not observed. 
The spurious phase is also present in microscopic Hartree Fock calculations (S.I.), and can be traced to the artificial SU(4) symmetry of interactions within these models. 
More accurately, the internal symmetry group of rhombohedral trilayer graphene consists of $SU(2)$ spin conservation, charge conservation, time reversal, and the lattice symmetries.  
Within this lower symmetry group, a variety of interactions that are anisotropic within the spin- and valley-space are allowed, particularly Hund's-type couplings which favor phases with particular broken spin and/or valley symmetries (see S. I.).  
This problem has been considered in the context of spontaneous symmetry breaking 
in graphene quantum Hall ferromagnets\cite{alicea_interplay_2007,kharitonov_phase_2012}, taking the form of an intervalley spin exchange coupling that favors the formation of a canted antiferromagnetic state at charge neutrality. 
Motivated by this observation, we introduce a flavor anisotropy of the form $\Phi'/A=JA_{\rm u.c.}\left(n_{K\uparrow}-n_{K\downarrow}\right)\left(n_{K'\uparrow}-n_{K'\downarrow}\right)$. 
For $J>0$, this term favors opposite spin polarizations in the two valleys, as is thought to occur in graphene quantum Hall ferromagnets, while for $J<0$ this term corresponds to a Hund's coupling and favors valley unpolarized, spin ferromagnetic ground states.  The phase diagram including this term---which is independent of the sign of $J$, is shown in Fig.~\ref{fig2}c-d for $U=30$\,eV, with $|J|/U=0.3$. 

To constrain the precise broken symmetries in the half- and quarter-metal phases,  we study the evolution of the phase transitions in an in-plane magnetic field, which couples primarily to the electron spin through the Zeeman effect. The resulting change in density at which a given transition occurs is directly proportional to the difference in Zeeman energy between the two competing phases, making tilted fields a sensitive probe of relative spin polarization. 
As shown in Fig.~\ref{fig2}e,  Zeeman energy favors the half-metal over the fully symmetric---and necessarily spin unpolarized---state. Moreover, the phase transition density shows a cusp at $B_\parallel=0$, implying an energy difference that is linear in $B_\parallel$ as expected for a ferromagnetic half metal with a divergent spin susceptibility at $B=0$. In contrast, the transition between half- and quarter metal is unaffected by $B_\parallel$ as expected for identical--and presumably full---spin polarization in both phases. Measurements of the Hall effect (Figs.~\ref{fig2}f and \ref{fig:S:anomalousHall}) show anomalous Hall effect in the quarter-metal phase but no corresponding effect in the half-metal.  This is expected due to the contrasting Berry curvatures in the two valleys, which cancel for valley unpolarized states but may give rise to an intrinsic anomalous Hall effect for valley polarized states\cite{xiao_berry_2010}.
Taken together, we conclude that the quarter metal is spin- and valley polarized while the half-metal is spin polarized but valley unpolarized (Fig. \ref{fig2}g). Interestingly, this implies a ferromagnetic Hund's coupling ($J<0$), in contrast to the the quantum Hall ferromagnet in mono- and bilayer-graphene\cite{dean_fractional_2020}. 

\subsection{Ferromagnetism and Fermi surface topology in the valence band}

\begin{figure*}[ht!]
\centering
\includegraphics[width=\textwidth]{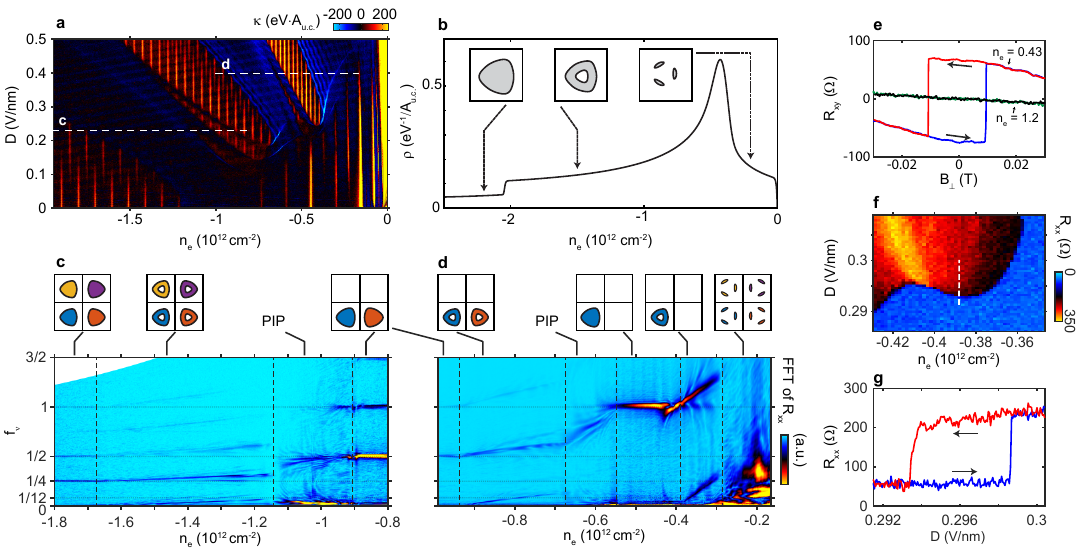}
\caption{\textbf{Ferromagnetism and Fermi surface topology in the valence band},
\textbf{a}, $\kappa$ measured at $B_{\perp}=1$\,T in the valence band.
\textbf{b}, Density of states as a function of carrier density in the valence band for $\Delta_1=20$\,meV calculated from the continuum model for a single isospin flavor. 
Insets: Calculated Fermi contours at the indicated values of $n_e$, plotted on the interval $(-0.05,0.05) a_0^{-1}$ for both $k_x$ and $k_y$. 
\textbf{c}, Fourier transform of quantum oscillations in resistivity measured at $D=0.23\,$V/nm along the the dashed line indicated in panel a (see Methods ). 
Insets show schematic depictions of Fermi contours in density domains distinguished by their quantum oscillations.  
\textbf{d}, 
As in panel c, but for $D=0.40\,$V/nm.  
\textbf{e}, 
Low magnetic field Hall resistivity measured at $D=0.43$V/nm at two values of $n_e$. Hysteretic anomalous Hall effect is observed in the valley-polarized quarter metal, but not in the valley unpolarized half metal.
\textbf{f},
Detail of the phase boundary between the quarter metal and symmetric low-density state. 
\textbf{g}, Resistivity measured as a function of $D$ across this boundary.  The transition is strongly first order, showing hysteresis as a function of applied gate voltages. 
}\label{fig3}
\end{figure*}

As compared to electron doping, hole-doped rhombohedral trilayer shows a considerably more complex phase diagram, as is seen in the $B_\perp=1\,{\rm T}$ magnetocapacitance data shown in Fig.~\ref{fig3}a. The contrast between the phase diagrams of the valence and conduction bands can be related to the single particle band structure, which differs markedly between the two.  Most importantly, in the valence band the density of states diverges at a finite density $n_{\rm e}^*\approx -5\times10^{11}$cm$^{-2}$, which corresponds to the merger of three disjoint Fermi pockets at low hole density into a single annular Fermi surface.  At still higher $|n_{\rm e}|$, the small electron pocket centered at each corner of the Brillouin zone disappears, leading to a step discontinuity (see Fig.~\ref{fig3}b). 
As a result, density-driven phase transitions in the valence band may be of several general types.  First, as in the conduction band isospin symmetries may break, reducing the  degeneracy of the Fermi surface.  In addition, Lifshitz transitions in the topology of the Fermi surface, which are already evident in the single particle band structure, may occur.  Finally, the nonmonotonic dependence of density of states on $n_e$ may favor states with partial isospin polarization, analogous to conventional ferromagnets, and allowing for Lifshitz transitions of a second type in which new Fermi surfaces are nucleated in previously unoccuppied spin/valley flavors.  

To disentangle these phases experimentally, we measure the resistivity as a function of perpendicular magnetic field and $n_e$ and plot the Fourier transform of $R_{xx}(1/B)$ (Figs. \ref{fig3}c-d, see also Fig. \ref{fig:S:sdh_p_side}) with frequencies normalized to that corresponding to the total carrier density. Peak position thus indicates fractional share of the total electrons enclosed by a given Fermi contour. At the highest values of $|n_e|$, a single peak (and its harmonics) is visible at $f_\nu=.25$, consistent with four fermi surfaces each enclosing an equal share of the total density.  As $n_e$ crosses the threshold of $n_e\approx 1.7\times 10^{12}\,$cm$^{-2}$ in Fig. \ref{fig3}c (corresponding to step in compressibility highlighted in Figs. \ref{fig1}d-e) the frequency of the quarter-density peak begins to grow and a second peak emerges at low frequency.  We interpret this as indicating a Lifshitz transition where a small electron Fermi surface is nucleated in the middle of the (now annular) Fermi sea, precisely as predicted by the single particle band structure.

Upon further lowering of $|n_e|$ towards zero, a sudden transition in the quantum oscillations is observed near $n_e\approx -1.15 \times 10^{-12}$ in Fig. \ref{fig3}c.  This threshold corresponds to a subtle but visible low $\kappa$ feature in Fig. \ref{fig1}d.  On the low $|n_e|$ side of the transition, the oscillation frequencies are less well defined, but but show spectral weight concentrated most prominently at $f_\nu$ slightly less than 0.5 and at very low frequencies.  These features are consistent with a Stoner-type transition to a partially isospin polarized (PIP) phase, with majority and minority charge carriers in two distinct pairs of isospin flavors. These contours continuously evolve until the $f_\nu$ of the high frequency peak converges to 0.5, whereupon the low frequency peak disappears, consistent with a Lifshitz transition from the PIP phase into a half-metal.  Remarkably, this pattern repeats itself as the density is lowered further, as shown in Fig. \ref{fig3}d: the simple half-metal transitions into an half-metal with an annular fermi sea, then to a PIP phase with one majority and one minority flavor, then into a simple quarter metal and then into an annular quarter metal before isospin symmetry is restored at the lowest densities. At these very low densities, each isospin flavor hosts three Fermi pockets, leading to observed oscillation frequencies near $f_\nu=1/12$. 

As for the electron side, many of these transitions show characteristic $B_\parallel$-dependence (Fig. \ref{fig:S:cpbp}) and anomalous Hall effect (Fig. \ref{fig:S:anomalousHall}) that allow us to confirm the spin- and valley-polarizations of the half- and quarter-metal states, which we find to be similar to those on the electron side.  We note, however, that transitions involving the PIP phases do not generally show simple linear-in-$B_\parallel$ behavior; combined with the complexity of the finite $B_\perp$ magnetoresistance suggests that these domains may harbor multiple PIP phases.

\subsection{Effect of a moir\'e superlattice potential}

\begin{figure}[ht!]
\centering
\includegraphics[width=\columnwidth]{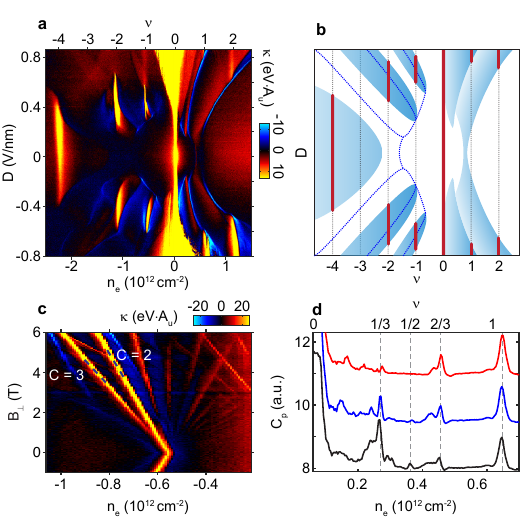}
\caption{\textbf{Effect of a moir\'e superlattice potential} 
\textbf{a}, $\kappa$ measured in sample B where the lattice of the rhombohedral trilayer graphene is aligned to that belonging to one of the encapsulating hBN crystals.
\textbf{b}, Schematic illustration of the formation of insulating states. Dashed lines indicate Lifshitz transitions described in Fig. \ref{fig3} and blue regions indicate domains of half- and quarter metal states in the absence of the moir\'e potential. 
Red lines represent the insulating states  observed in the moir\'{e} superlattice sample, which occur whenever the moir\'e filling is divisible by the degeneracy of the underlying metal. 
\textbf{c}, $\kappa$ in samples B measured at $D = -0.56$\,V/nm as a function of $n_{\rm e}$ and $B_\perp$.
\textbf{d}, Penetration field capacitance data as a function of $n_{\rm e}$ at $D = -0.82$\,V/nm (black), $-0.86$\,V/nm (blue) and $-0.89$\,V/nm (red). The curves are offset for clarity. Filling factors of some incompressible states are marked on the top axis.
}\label{fig4}
\end{figure}

Experimental indications of strong interactions in intrinsic rhombohedral graphene have previously been restricted to the very low density ($|n_{\rm e}|<10^{11} \rm{cm^{-2}}$) regime\cite{shi_electronic_2020,lee_gate_2019}.
Recently, however, manifestations of strong interaction have emerged in rhombohedral trilayers aligned to hexagonal boron nitride at densities comparable to the phase transitions reported here\cite{chen_evidence_2019,chen_signatures_2019,chen_tunable_2020}.  
In these devices, insulating states have been observed at filling $\nu=-1$ and $-2$ of the superlattice unit cell, including an incipient Chern insulator at $\nu=-1$. These experimental findings were interpreted as resulting from polarizing an emergent flat band into one or more valley and spin flavors.  
It is interesting to reexamine this picture in light of our finding that rhombohedral graphite spontaneously breaks spin and valley symmetry in the absence of a moir\'e potential, which amounts to quantifying the difference between moir\'e superlattice and non-moir\'e superlattice trilayer devices. 

We address this question directly by measuring the inverse compressibility data in a  rhombohedral trilayer device aligned to one of the encapsulating hexagonal boron nitride layers (Fig.~\ref{fig4}a) but that has otherwise identical geometry to the unaligned device presented in Figs.~\ref{fig1}-\ref{fig3}. 
We find the negative compressibility features associated with Stoner transitions nearly unchanged in the aligned, moir\'e device.  The primary difference is the appearance of incompressible states at commensurate fillings $\nu=\pm1,\pm2$ of the moir\'e unit cell.
The relationship between these insulators and the underlying symmetry breaking in non-moir\'e devices is depicted schematically in Fig.~\ref{fig4}b, where we overlay the phase boundaries measured in intrinsic trilayers with the domain of stability of the commensurate, incompressible states in moir\'e patterned trilayers.  Evidently, incompressible states emerge whenever the superlattice filling is divisible by the degeneracy of the Fermi surface at the same $n_{\rm e}$ and $D$ absent a moir\'e. The effect of the moir\'e can thus be understood as a perturbation that does not qualitatively alter the correlated electron physics already present in the parent trilayer. 

In crystalline systems incompressible gapped states can only occur at commensurate fillings of the lattice.  The moir\'e superlattice is qualitatively important in that it allows for gapped states at carrier densities that can be reached by electrostatics gates, and we observe several classes of commensurate gapped states driven by electron interaction in our high quality samples.  Gapped states are classified by two quantum numbers, $s$ and $t$, which respectively encode the number of electrons per lattice site and the Chern number, which is linked to the quantized Hall conductivity. We classify gaps by the resulting trajectories in the $n_{\rm e}$-$B_\perp$ plane, $\nu=t n_\phi+s$, where $n_\phi$ is the number of magnetic flux quanta per unit cell. 
Consistent with prior work, we find that commensurate insulators at $\nu=-1$ and $\nu=-2$ states are topologically trivial for $D>0$, with $(s,t)=(-1,0)$ and $(-2,0)$, respectively (see Fig.~\ref{fig:S:chern0supp}).  In contrast, the  $\nu=-1$ insulators is nontrivial for $D<0$ (Fig.~\ref{fig4}c).
Our high resolution data allow us to observe a close competition between robust $t=-2$ and $t=-3$ Chern insulators for $s=-1$.  
At high magnetic field, these states occur at different densities, and high inverse compressibility peaks are observed corresponding to both trajectories.  As $B_\perp$ tends to zero and the states converge to the same density, the $t=-2$ state wins the energetic competition, consistent with transport data (see Fig.~\ref{fig:S:chern0supp} and Ref.~\onlinecite{chen_tunable_2020}).  

As shown in Figure \ref{fig4}d, we also observe a number of features at \textit{fractional} filling of the moir\'e superlattice bands. These states are all found to have $t=0$ in the low $B_\perp$ limit, and occur at $\nu=s=1/3,1/2,2/3...$.  The regime where these states are observed corresponds within single particle band structure models to a regime where an unusually flat topologically trivial flat band is partially filled\cite{zhang_nearly_2019}.  We interpret them as generalized Wigner crystals, in which electron repulsion leads to commensurate filling of the moir\'e potential, breaking the superlattice symmetry.  Similar states have been reported in transition metal dichalcogenide heterobilayers\cite{regan_mott_2020,xu_correlated_2020}.  

\section{Discussion}

It is interesting to compare the itinerant ferromagnetism revealed in rhombohedral trilayer graphene with other physical realizations of ferromagnetism.  Unlike in transition metals, ferromagnetism in trilayer graphene does not originate from tightly localized atomic orbitals, but rather is intrinsic to the  itinerant electrons themselves. Rhombohedral trilayer graphene may also be contrasted with low density semiconductors\cite{pudalov_thermodynamic_2015,hossain_observation_2020-1}.  While superficially similar, signatures of ferromagnetism in these systems are observed in an insulating regime, where electron wave functions are localized either by disorder or Wigner crystallization.  

Despite the lack of an applied magnetic field, the underlying physics of rhombohedral trilayer graphene closely resembles that of a quantum Hall ferromagnets. Indeed, the sharp peak in the density of states associated with the van Hove singularity plays a similar role to a that of a Landau level: high density of states allows a large gain of exchange energy by isospin polarization, at a small cost in kinetic energy. As in a Landau level, the electronic wave functions corresponding to the high density of states region are strongly overlapping, enhancing the exchange coupling. 

These considerations apply equally well to isospin magnetism in moir\'e heterostructures, which is thought to underpin the correlated physics observed in these systems\cite{balents_superconductivity_2020}. 
From this point of view, it is the high density of states of the flat bands, and not their isolation from high energy dispersive bands, that plays the central role in the ferromagnetic order. As demonstrated here, the primary role of the moir\'e potential is to enable gapped states at finite density. In this light, our results suggest a new design space for van der Waals heterostructures, based on gate tunable isospin magnetism as a building block that is both strongly correlated but also well understood. 

\section{Methods}
The trilayer graphene and hBN flakes were prepared by mechanical exfoliation of bulk crystals. The rhombohedral domains of trilayer graphene flakes were detected by a Horiba T64000 Raman spectrometer with a 488nm mixed gas Ar/Kr ion laser beam. The rhombohedral domains were subsequently isolated using a Dimension Icon 3100 atomic force microscope. The Van der Waals heterostructures were fabricated following a dry transfer procedure. A special stacking order was followed to minimize the mechanical stretching of rhombohedral trilayer graphene. The details are described in Fig.~\ref{fig:S:fab}.

All electronic measurements were performed in dilution refrigerators equipped with a superconducting magnet. And capacitance data for the unaligned sample A are largely symmetric under $D\rightarrow-D$; however, for $n_e\times D>0$, a high resistance single-gated region is introduced to the contact area where only one of the gates acts on the channel.  We thus focus on the  $n_e\times D<0$ quadrants in the bulk of our analysis, with data over the dull $n_e$, $D$ range shown in Fig. \ref{fig:S:cap_full}.  

Penetration field capacitance was measured using a capacitance bridge circuit with an  an FHX15X high electron mobility transistor serving as an in situ impedance transformer\cite{zibrov_tunable_2017}. An excitation frequency of 10245.12Hz was used to obtain the data in Fig. \ref{fig4}d. The rest of the capacitance data were measured at 54245.12Hz. 
The quantity directly measured is $M=(c_{\rm p}+c_{\rm parasitic})/c_{\rm ref}$, where $c_{\rm p}$ is the capacitance between the top and bottom gate, $c_{\rm ref}$ is the capacitance of the reference capacitor, $c_{\rm parasitic}$ is the parasitic capacitance of the instrument. 
The inverse compressibility is related to $c_{\rm p}$ by 
$c_{\rm p}=\frac{c_{\rm t}c_{\rm b}}{c_{\rm t}+c_{\rm b}+\kappa^{-1}}\approx\kappa c_{\rm t}c_{\rm b}$\cite{zibrov_tunable_2017},
where $c_{\rm t(b)}$ is the geometric capacitance between the top (bottom) gate and the trilayer graphene.
To obtain $\kappa$, $M$ is measured at two extremes, denoted $M_{\infty}$ and $M_0$.  $M_{\infty}$ corresponds to when the trilayer graphene is a good metal, and is achieved by applying a large out-of-plane magnetic field and tuning the Fermi level within a partially filled Landau level. $M_0$ corresponds to when the trilayer is incompressible, which can be achieved by applying a large displacement field $D$ while keeping the carrier density $n_{\rm e}=0$. 
In the former case $c_{\rm p}=0$, therefore $M_{\infty}=c_{\rm parasitic}/c_{\rm ref}$. 
In the later case, $c_{\rm p}=(M_{0}-M_{\infty})c_{\rm ref}=\frac{c_{\rm t}c_{\rm b}}{c_{\rm t}+c_{\rm b}}$. From the equations above, $\kappa = \frac{1}{2c}\frac{M-M_{\infty}}{M_{0}-M{\infty}}$,
where the averaged geometric capacitance $c = (c_{\rm t}+c_{\rm b})/2$ can be obtained by linear fitting of the carrier density $n_{\rm e}$, which is known for fixed Landau level filling providing a calibration standard.  

Transport measurement was performed using a lock-in amplifier. The frequency was chosen between 17.777Hz to 42.5Hz to minimize the noise. A series of cryogenic high-pass filters were applied to reduce the electron temperature.

To analyze the magnetoresistance oscillations, a fifth-order polynomial fit is subtracted from the $R_{xx}(B_\perp)$ data.  The data is then interpolated to produce an even grid as a function of $1/B_\perp$.  Fourier transforms are computed over a range of $B_\perp\in (.02T,.33T)$ for Fig. \ref{fig2}b, $(0.3T,1T)$ for Fig. \ref{fig3}c and $(0.02T,1T$ for Fig. \ref{fig3}d. The lower bound of $B_{\perp}$ is chosen by the lowest $B_\perp$ where oscillations are visible, and the upper bound is chosen to avoid obvious $B_\perp$-induced phase transitions.  Raw magnetoresistance data for the Fourier transforms shown in the main text are presented in Fig. \ref{fig:S:sdh_n_side} and Fig. \ref{fig:S:sdh_p_side}.

\section*{acknowledgments}
The authors acknowledge discussions with A. Macdonald, L. Fu, F. Wang and M. Zaletel. 
AFY acknowledges support of the National Science Foundation under DMR-1654186, and the Gordon and Betty Moore Foundation under award GBMF9471. 
The authors acknowledge the use of the research facilities within the California NanoSystems Institute, supported by the University of California, Santa Barbara and the University of California, Office of the President.
K.W. and T.T. acknowledge support from the Elemental Strategy Initiative
conducted by the MEXT, Japan, Grant Number JPMXP0112101001 and JSPS
KAKENHI, Grant Number JP20H00354. 
EB and TH were supported by the European Research Council (ERC) under grant HQMAT (Grant Agreement No.~817799).  
A.G. acknowledges support by the European Unions Horizon 2020 research and innovation program under the Marie Sklodowska-Curie Grant Agreement No.~754411.

\section*{Author contributions}
 HZ and TX fabricated the device with assistance from EMS and JE.  
 HZ performed the measurements, advised by AFY.  
 KW and TT grew the hexagonal boron nitride crystals.  
 AG, TR, EB, and MS contributed to the theoretical interpretation and performed the numerical simulations.  
 HZ, AG, MS, EB, and AFY wrote the manuscript with input from all authors.

\normalem
\let\oldaddcontentsline\addcontentsline
\renewcommand{\addcontentsline}[3]{}
\bibliographystyle{custom}
\bibliography{references}
\let\addcontentsline\oldaddcontentsline

\renewcommand\thefigure{S\arabic{figure}}
\setcounter{figure}{0}
\begin{figure*}
\centering
\includegraphics[width=180mm]{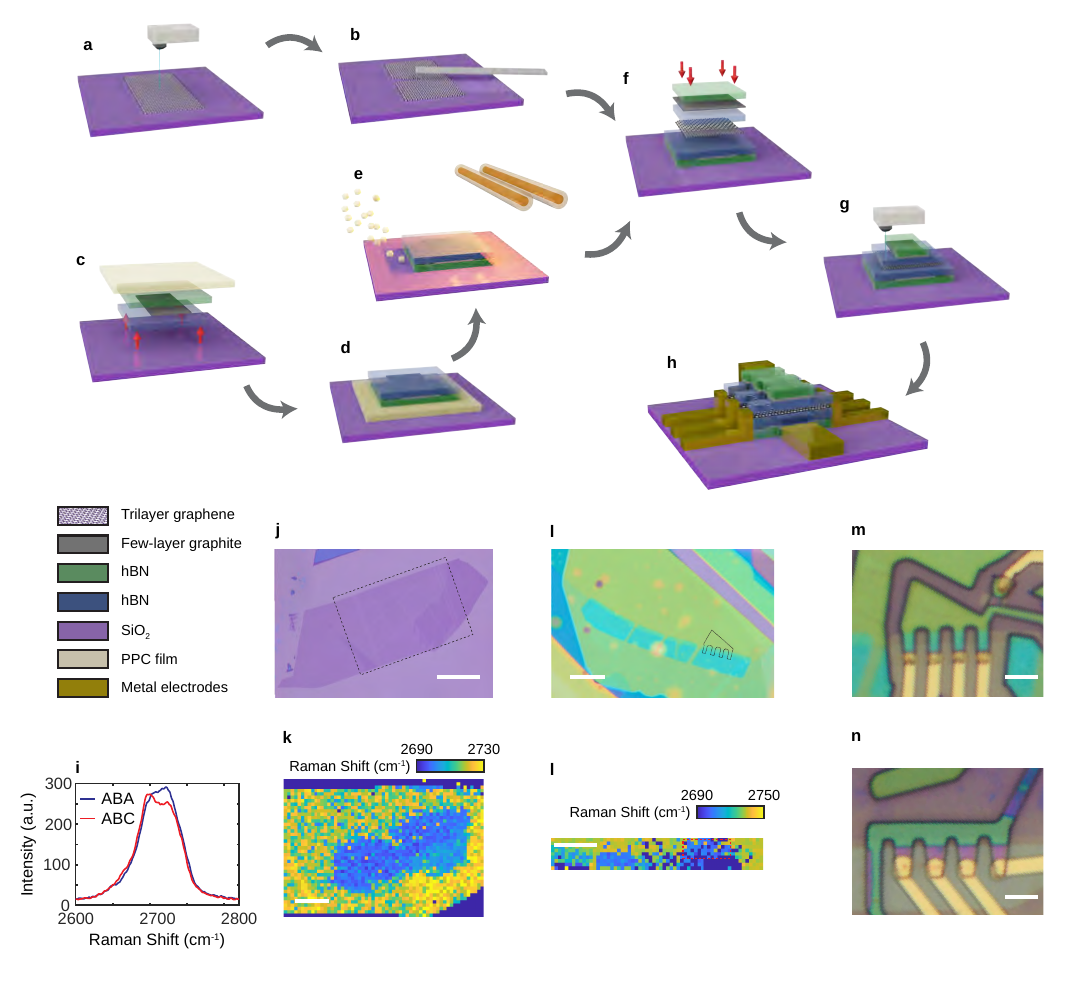}
\caption{\textbf{Sample fabrication procedure.}
\textbf{a}, ABC-stacked domains in mechanically exfoliated trilayer graphene flakes are identified by taking the Raman spectra and extracting the peak maximum corresponding to the 2D mode\cite{cong_raman_2011}. 
\textbf{b}, ABC-stacked domains are isolated using atomic force microscope based anodic oxidation lithography\cite{li_electrode-free_2018}. 
\textbf{c}, The lower part of the heterostructure is assembled on a polypropylene carbonate (PPC) film which is then 
\textbf{d}, flipped as it is deposited onto the target substrate\cite{polshyn_quantitative_2018}. 
\textbf{e},The sample is then vacuum annealed at 375$^\circ$ C to remove the PPC film under the heterostructure.
\textbf{f}, The upper part of the heterostructure, which contains the top graphite gate, trilayer graphene and hBN, is assembled separately and deposited onto the lower part of the heterostructure. 
\textbf{g}, The top hBN and top graphite gate are etched with XeF$_2$ followed by O$_2$ plasma to open windows on the heterostructure, allowing the stacking order to be confirmed after the manipulations of step f. 
\textbf{h}, The heterostructure is etched with CHF$_3$ and O$_2$ plasma and metal is deposited to form electrical contacts.
\textbf{i}, Typical Raman spectra of ABA- and ABC-stacked trilayer graphene, centered on the 2D mode.
\textbf{j}, Optical micrograph of the trilayer graphene flake used to fabricate Sample A. Scale bar represents 20 $\mu$m.
\textbf{k}, Raman spectrum map of the trilayer graphene flake in panel j. The color represents the peak position of the 2D mode. 
The scan range is indicated in black dashed line in panel j. The scale bar represent 10$\mu$m.
\textbf{l}, Optical micrograph of partially processed Sample A. The cyan regions are where the top graphite gate and the hBN on top of it has been etched. Since the bottom gate does not overlap with the etched window, this allows inspection with Raman spectroscopy of the stacking order of the trilayer graphene. The sale bar represents 10$\mu$m. The rough location of the actual device is indicated by black dashed line.
\textbf{m}, Raman spectrum map of the partially processed Sample A. The region surrounded by a red boundary box remains in ABC-stacking order, which later became the active device region for sample A.
\textbf{n}, Optical micrograph of Sample A after fabrication. Scale bar represents 3$\mu$m.
\textbf{o}, Optical micrograph of Sample B. Scale bar represents 3$\mu$m.
}\label{fig:S:fab}
\end{figure*}

\begin{figure*}
\centering
\includegraphics[width=\textwidth]{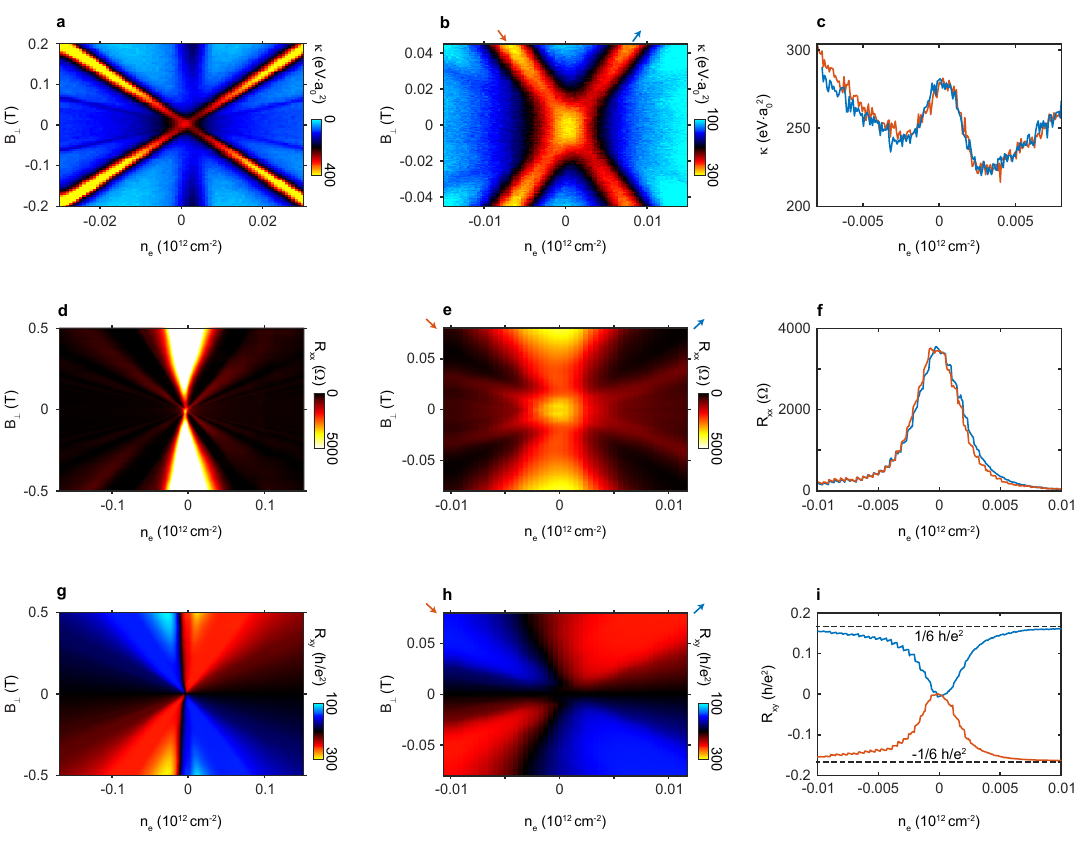}
\caption{\textbf{Measurements at low carrier density and displacement field.}
\textbf{a}, Inverse compressibility versus carrier density and out-of-plane magnetic field near the charge neutrality point at $D=0$.
\textbf{b}, Zoom-in of a.
\textbf{c}, Line-cuts of b along the direction indicated by the arrows in b.
\textbf{d}, $R_{xx}$ versus carrier density and out-of-plane magnetic field near the charge neutrality point at $D=0$.
\textbf{b}, Zoom-in of d.
\textbf{c}, Line-cuts of e along the direction indicated by the arrows in e.
\textbf{g}, $R_{xy}$ versus carrier density and out-of-plane magnetic field near the charge neutrality point at $D=0$.
\textbf{h}, Zoom-in of g.
\textbf{i}, Line-cuts of h along the direction indicated by the arrows in h.
}\label{fig:S:transport_cnp}
\end{figure*}

\begin{figure*}
\centering
\includegraphics[width=\textwidth]{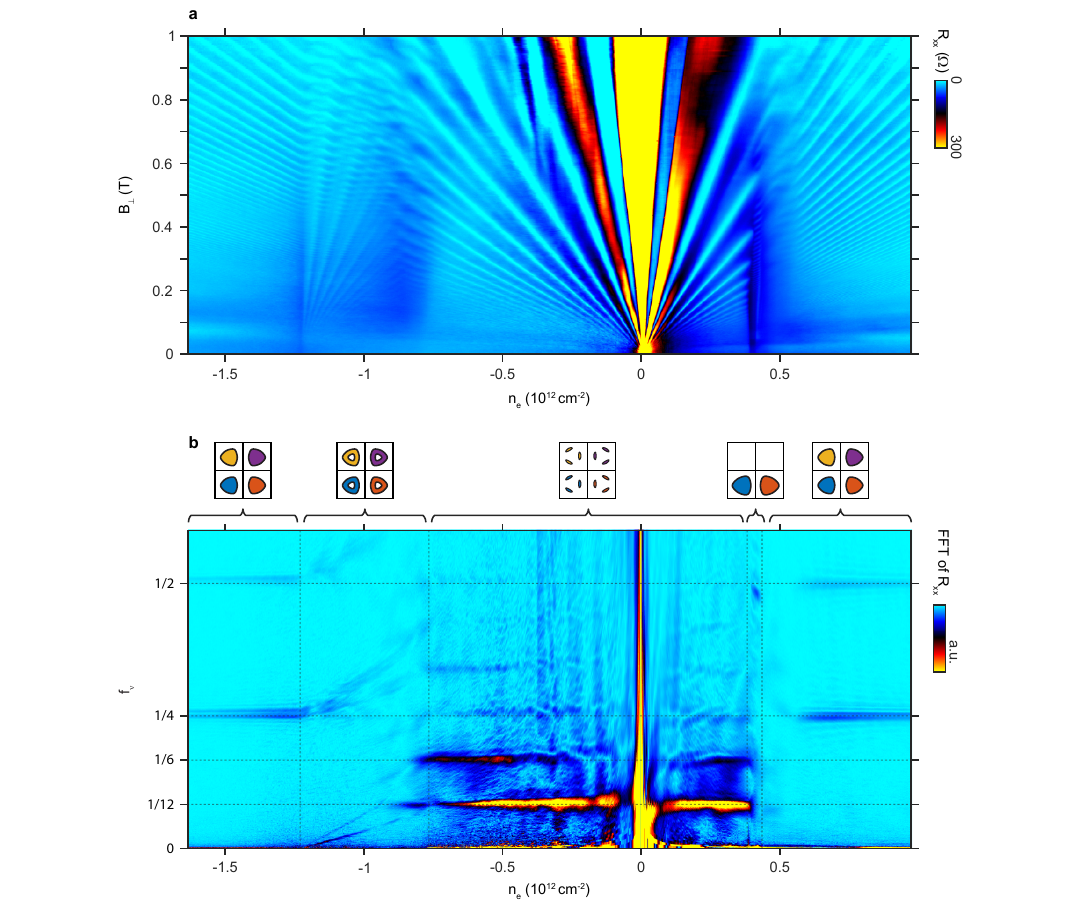}
\caption{\textbf{Shubnikov de Haas oscillation at D = 0 in Sample A.}
\textbf{a}, $R_{xx}$ vs $n_{\rm e}$ and $B_{\perp}$ measured at $D=$0.
\textbf{b}, Fast Fourier transform of data in a, the range of $B_\perp$ chosen is 0.02T to 1T. The multiple phases are schematically represented by the Fermi contours on the top.
}\label{fig:S:FFT_at_p0=0}
\end{figure*}

\begin{figure*}
\centering
\includegraphics[width=\textwidth]{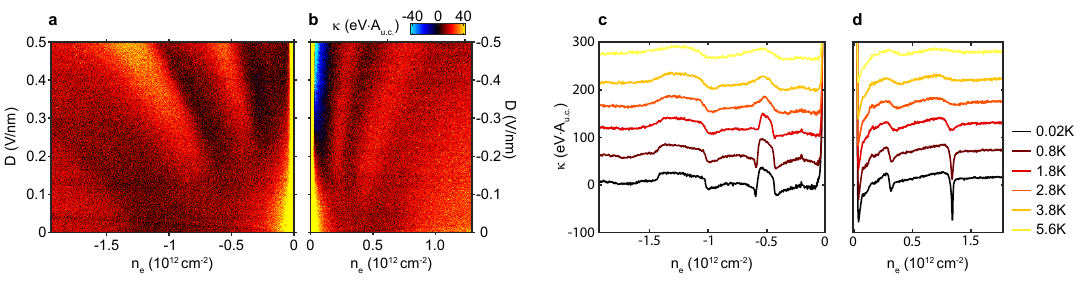}
\caption{\textbf{Temperature dependence of $\kappa$ in Sample A.}
\textbf{a-b}, Inverse compressibility as function of carrier density and displacement measured at $T=5.6$K.
\textbf{c-d}, Inverse compressibility as a function of carrier density measured measured at various temperatures at $D=0.46$\,V/nm (c) and $-0.46$\,V/nm (d). Each curve is offset by 50 eV$ \cdot A_{\rm a.u.}$ for clarity.
}\label{fig:S:capacitance}
\end{figure*}

\begin{figure*}
\centering
\includegraphics[width=\textwidth]{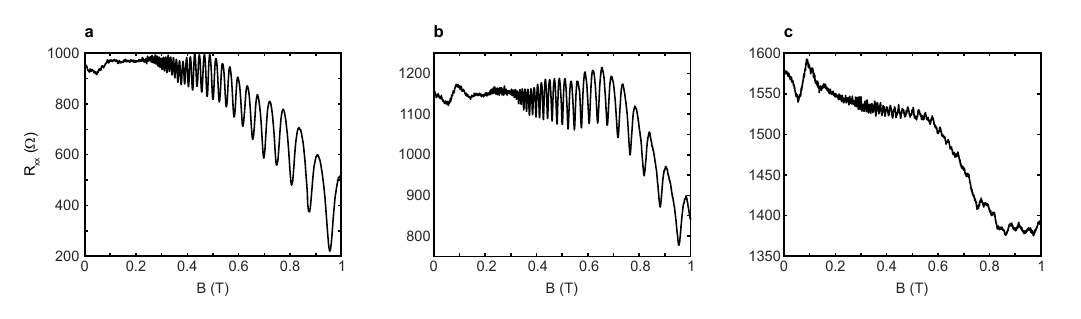}
\caption{\textbf{Shubnikov de Haas oscillations in Sample A with positive electron density.} The measurement are performed at $D=$-0.43V/nm at $n_{\rm e}=0.27 \times \rm{cm}^{-2}$ in a, $0.54 \times \rm{cm}^{-2}$ in b and $1.09 \times \rm{cm}^{-2}$ in c. The fast Fourier transform in Fig. \ref{fig2}b are calculated from these results.
}\label{fig:S:sdh_n_side}
\end{figure*}

\begin{figure*}
\centering
\includegraphics[width=\textwidth]{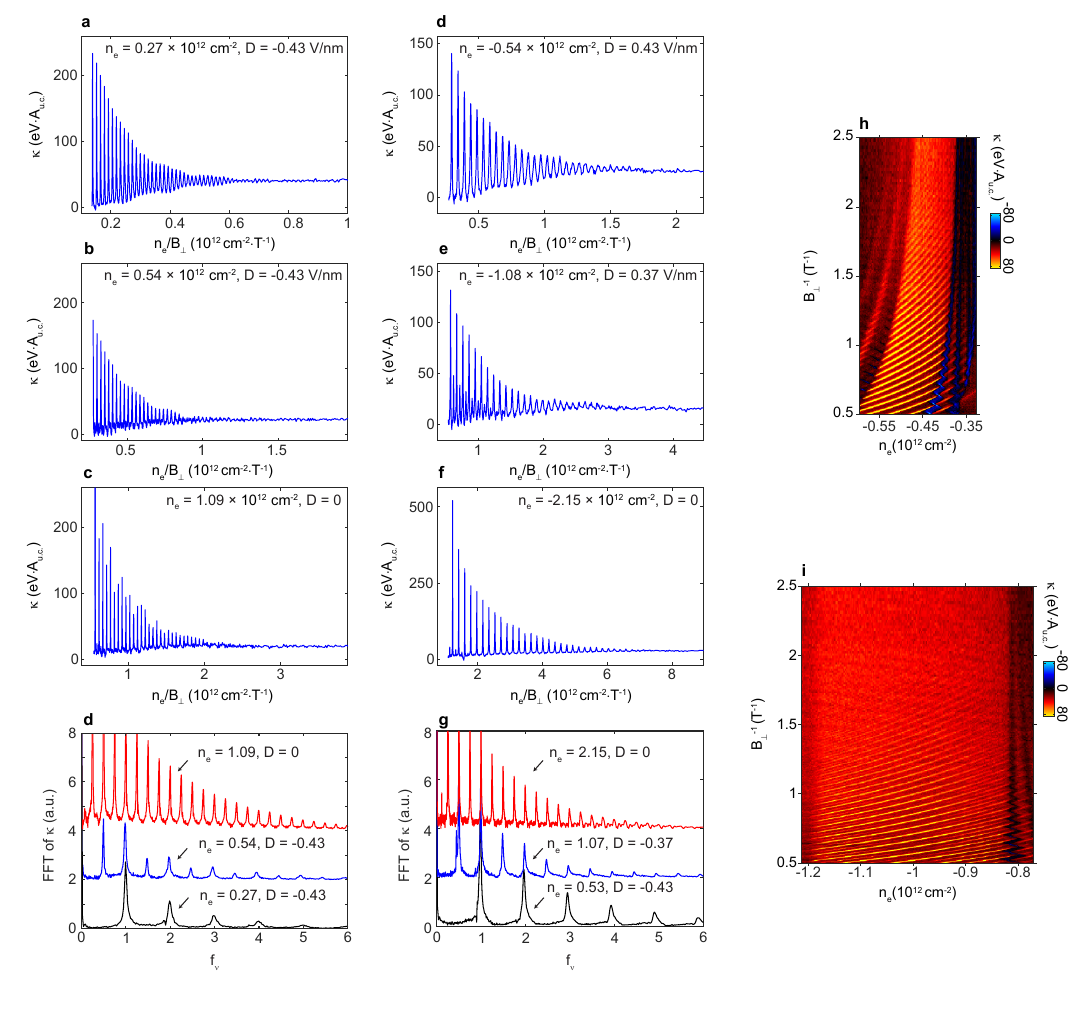}
\caption{\textbf{Quantum capacitance magneto-oscillations.}
\textbf{a}-\textbf{f}, Inverse compressibility as a function of the out-of-plane magnetic field at fixed carrier density and displacement fields. The results in Fig.\ref{fig2}b are generated by calculating the fast Fourier transform of these results.
\textbf{g} and \textbf{h}, Inverse compressibility versus the out-of-plane magnetic field and the carrier density at a fixed displacement field of 0.34\,V/nm. The range of carrier density are chosen to lie within the one-fold degenerate phase in panel g and two-fold degenerate phase in panel h; in both cases no change in the degeneracy is observed in the low-$B_\perp$ limit. 
}\label{fig:S:oscillation}
\end{figure*}

\begin{figure*}
\centering
\includegraphics[width=\textwidth]{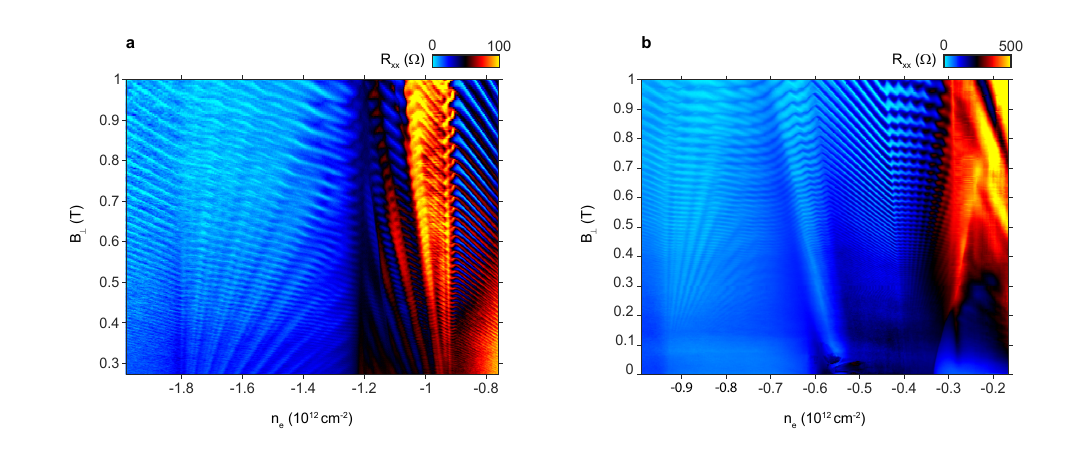}
\caption{\textbf{Shubnikov de Haas oscillations in Sample A with negative electron density.} Data in panel a is measured at $D=$0.23V/nm. Data in panel d is measured at $D=0.4V/nm$. The fast Fourier transform in Fig. \ref{fig3}c and d are calculated from these results.
}\label{fig:S:sdh_p_side}
\end{figure*}

\begin{figure*}
\centering
\includegraphics[width=\textwidth]{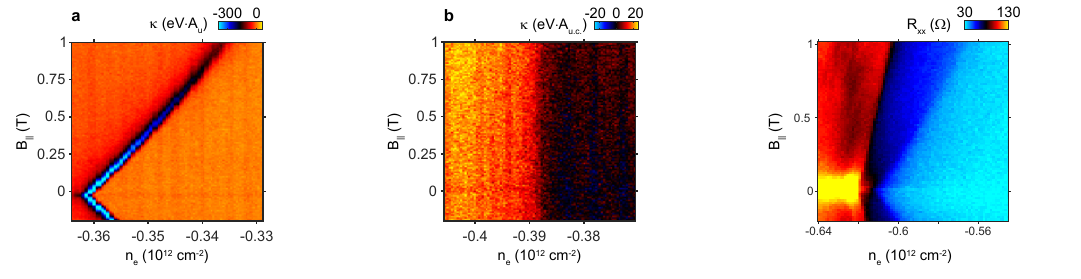}
\caption{\textbf{In-plane magnetic field dependence of the phase boundaries}. 
\textbf{a}, $\bm{\kappa}$ vs $\bm{n_{\rm e}}$ and $\bm{B_{\parallel}}$ at $D=$0.37V/nm, which covers  the phase boundary between a 4-fold degenerate phase and a 1-fold degenerate phase.
\textbf{b}, Same as a, measured at $D=0.33$V/nm, which covers a phase boundary between a 1-fold degenerate phase with a simple Fermi surface and a 1-fold degenerate phase with annular Fermi surface.
\textbf{c}, $\bm{R_{xx}}$ vs $\bm{n_{\rm e}}$ and $\bm{B_{\parallel}}$ at $D=$0.37V/nm, which covers a phase boundary between a 4-fold degenerate phase and a 2-fold degenerate phase.
}\label{fig:S:cpbp}
\end{figure*}

\begin{figure*}
\centering
\includegraphics[width=\textwidth]{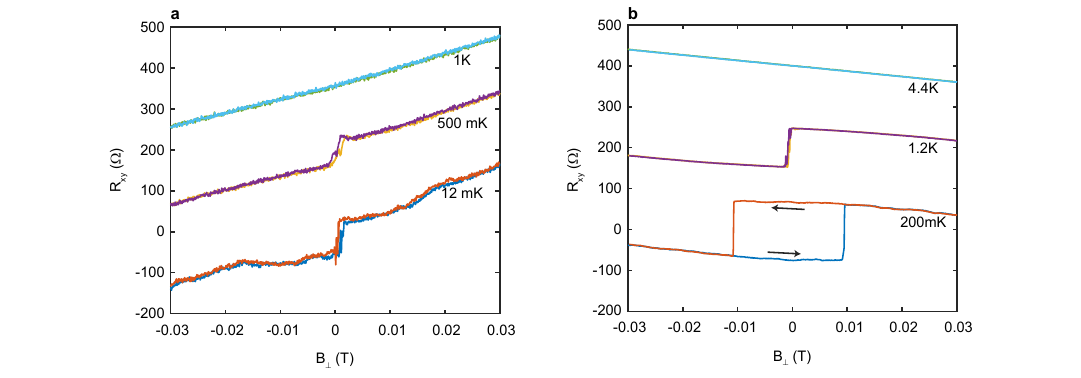}
\caption{\textbf{Anomalous Hall effect and Magnetic hysteresis.} 
\textbf{a}, Hall resistance $R_{xy}$ as a function of the out-of-plane magnetic field $B_{\perp}$ measured at $n_{\rm e}=0.19\times10^{12}{\rm cm}^{2}$, $D=-0.4$V/nm. 
\textbf{b}, Same measurement at $n_{\rm e}=0.43\times10^{12}{\rm cm}^{2}$, $D=0.38$V/nm with an 0.1T in-plane magnetic field applied.
The curves measured at different temperatures are offset by 200 $\Omega$ for clarity. The Hall resistance were obtained by measuring the four-terminal resistance in two configurations and applying the Onsager reciprocal relation\cite{serlin_intrinsic_2020}.
}\label{fig:S:anomalousHall}
\end{figure*}

\begin{figure*}
\centering
\includegraphics[width=\textwidth]{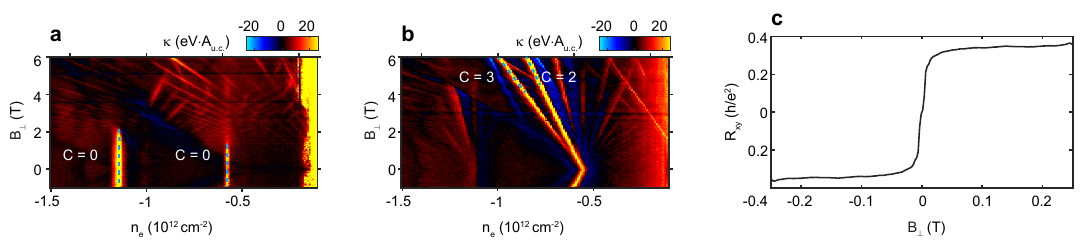}
\caption{\textbf{Magnetic field response of the insulating states in Sample B}
\textbf{a}, Inverse compressibility versus carrier density and out-of-plane magnetic field at $D = 0.52$\,V/nm measured in sample B.
\textbf{b}, same as c measured at D = -0.57 V/nm.
\textbf{c}, $R_{xy}$ versus $B_\perp$ measured at $n_{\rm e}=-0.52\times10^{12}$\,V/nm, $D=-0.47$\,V/nm.
}\label{fig:S:chern0supp}
\end{figure*}

\begin{figure*}
\centering
\includegraphics[width=\textwidth]{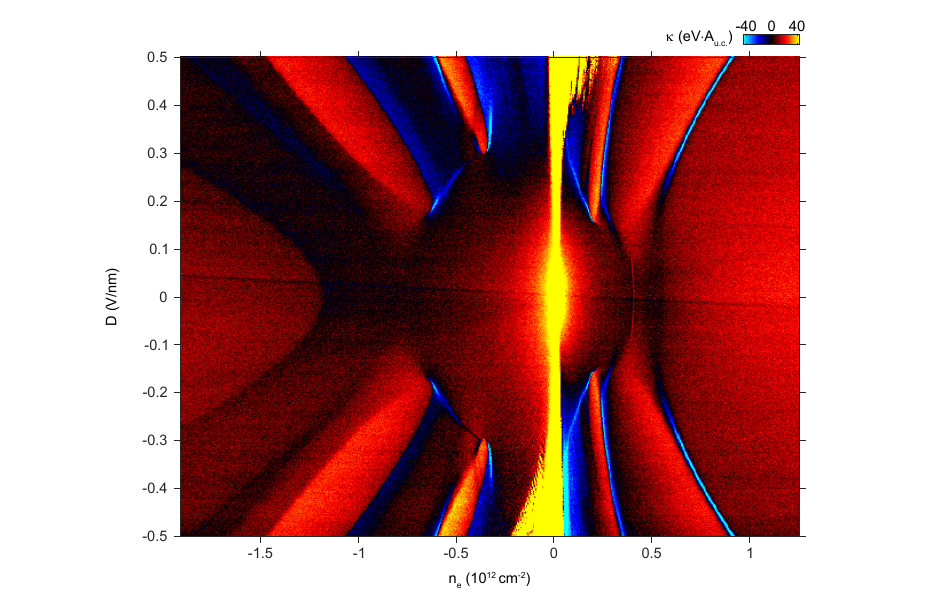}
\caption{\textbf{Extended $\bm{\kappa}$ vs $\bm{n_{\rm e}}$ and $\bm{D}$ data at $\bm{B=}$0.} The contact resistance increase at $n_{\rm e}>0$, $D>0$ and at $n_{\rm e}<0$, $D<$0 due to the formation of a pn-junction near the contact, producing the defect features near the charge neutrality point.
}\label{fig:S:cap_full}
\end{figure*}

\clearpage
\onecolumngrid
\begin{center}
\textbf{\large Supplementary information for ``Half and quarter metals in rhombohedral trilayer graphene'' }\\[5pt]
\begin{quote}
 {\small 
}
\end{quote}
\end{center}
\setcounter{equation}{0}
\setcounter{table}{0}
\setcounter{page}{1}
\setcounter{section}{0}
\makeatletter
\renewcommand{\theequation}{S\arabic{equation}}
\renewcommand{\thefigure}{S\arabic{figure}}
\renewcommand{\thepage}{S\arabic{page}}


\section{Theoretical model of rhombohedral trilayer graphene}\label{Sec:non-int}
\subsection{Band structure model}
The starting point for the theoretical models used in this work is a six-band continuum model for rhombohedral graphene adapted from Ref.~\onlinecite{zhang_band_2010}:
\begin{equation}
H_0 =
\left(\begin{matrix}
	\Delta_1+\Delta_2 +\delta &\frac12\gamma_2 &v_0 \pi^\dagger &v_4\pi^\dagger &v_3\pi &0 \\
\frac12\gamma_2	& \Delta_2-\Delta_1+\delta& 0&v_3 \pi^\dagger &v_4\pi & v_0\pi\\
v_0\pi	&0 &\Delta_1+\Delta_2 & \gamma_1&v_4 \pi^\dagger &0 \\
v_4\pi	&v_3\pi &\gamma_1 &-2\Delta_2 &v_0\pi^\dagger & v_4 \pi^\dagger\\
v_3\pi^\dagger	&v_4\pi^\dagger &v_4\pi & v_0\pi& -2\Delta_2&\gamma_1 \\
0	&v_0\pi^\dagger & 0& v_4\pi&\gamma_1 &\Delta_2-\Delta_1 \\     
\end{matrix}
\right),
\label{Ham6}
\end{equation}
where $\pi=\xi k_x+ik_y$ ($\xi=\pm$ corresponds to valleys $K$ and $K'$) and the Hamiltonian is written in the basis $(A_1,B_3,B_1,A_2,B_2,A_3)$, where $A_i$ and $B_i$ denote different sublattice sites on layer $i$.  
This Hamiltonian depends on band structure parameters $v_i=\sqrt{3}a_0\gamma_i/2$ with $i=0,3,4$ and bare hopping matrix elements $\gamma_1$ and $\gamma_2$, where
$a_0=2.46$\AA is the lattice constant of monolayer graphene. In addition, the parameter $\delta$ encodes an on-site potential which is only present at sites $A_1$ and $B_3$ since these two atoms do not have a neighbor on the middle layer. Finally, electrostatic potentials on different layers enter Eq.~(\ref{Ham6}) via $\Delta_{1,2}$. $\Delta_1$ corresponds to a potential difference between outer layers and is approximately proportional to the applied displacement field, while $\Delta_2$ is the potential difference between the middle layer compared to mean potential of the outer layers. 

\begin{figure*}
	\begin{center}
		\includegraphics[width=\columnwidth]{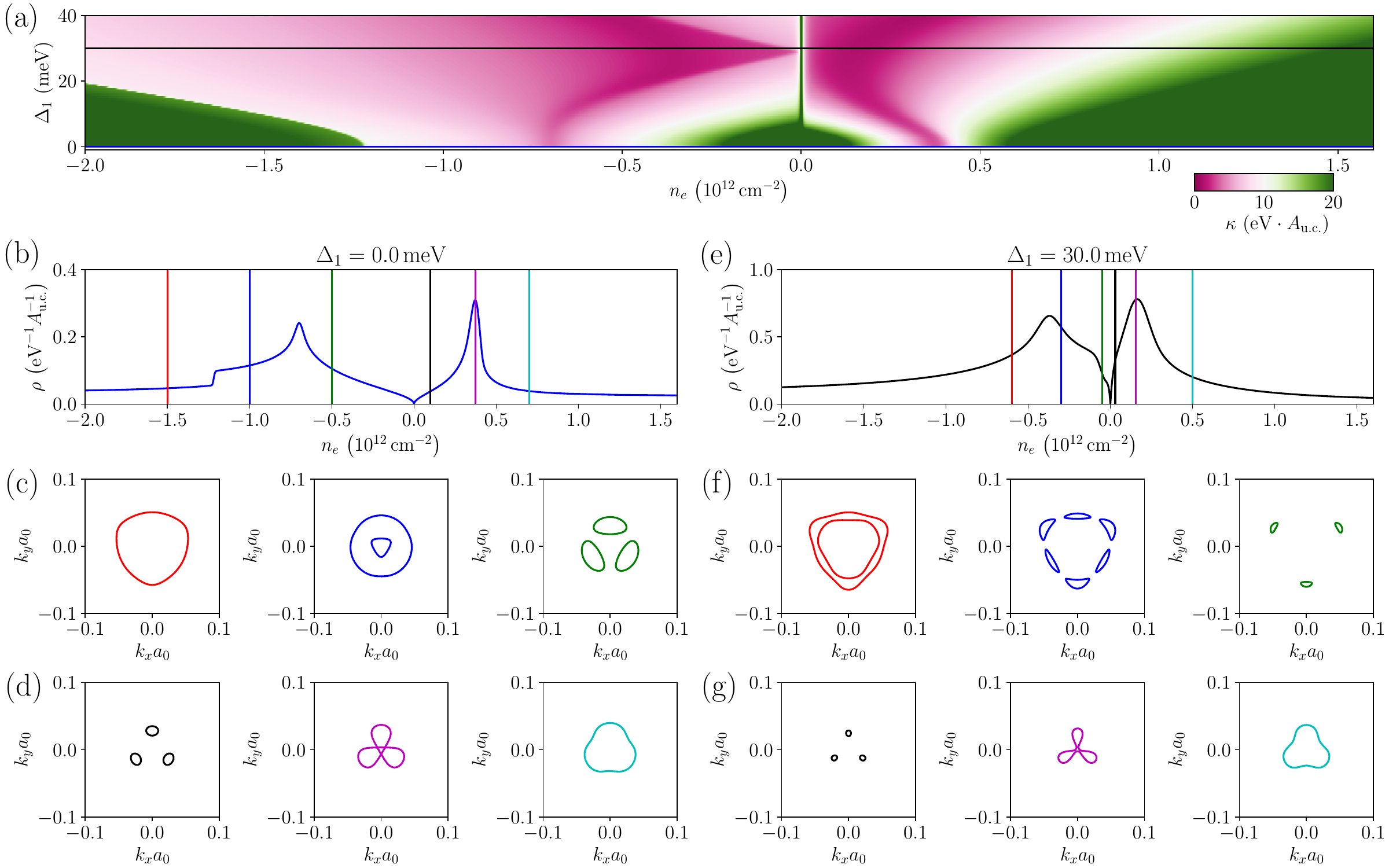} \\
		\caption{ \label{Fig:NonInteractingDOS}
			(a) The contour plot of inverse compressibility as a function of density and displacement field $\Delta_1$. (b) Density of states as a function of free carrier density at fixed $\Delta_1=0.0\,\mathrm{meV}$ [cut corresponding to blue line in (a)]. Fermi surfaces at six different fillings of valence and conduction bands are shown in panels (c) and (d) respectively. Panels (e)-(f) show similar data but for fixed value of $\Delta_1=30.0\,\mathrm{meV}$ [cut corresponding to black line in (a)]. The data shown in panels (a,b,e) is generated using a Fermi-Dirac distribution with finite temperature $T=0.025\,\mathrm{meV}$, leading to smoothing of the peaks related to vHs.}
	\end{center}
\end{figure*}

The values of tight-binding parameters $\gamma_{0,\ldots,4}$ and $\delta$ have been constrained in the literature from fits to experimental data~\cite{shi_electronic_2020,yin_high-magnetic-field_2019} and ab initio simulations~\cite{koshino_interlayer_2010,zhang_band_2010,jung_accurate_2014,chittari_gate-tunable_2019}. However, due to the large number of parameters and limited data availability, there exist significant discrepancies between these estimates. Since interactions play a crucial role in our experiment, fitting parameters using penetration field capacitance measurements, as some of us did previously in for Bernal-stacked trilayer graphene~\cite{zibrov_emergent_2018}, is also not feasible. 

\begin{table}[ht!]
\centering
\begin{tabular}{ccccccc}
\hline\hline
$\gamma_0$ & $\gamma_1$ &$\gamma_2$ & $\gamma_3$ &$\gamma_4$ &$\delta$ &$\Delta_2$ \\
\hline
$3.1^{\rm a}$ & $0.38^{\rm a}$ &  $-0.015$ & $-0.29^{\rm a}$ & $-0.141^{\rm a}$ & $-0.0105$ & $-0.0023$ \\
\hline\hline
\end{tabular}
\caption{Values of tight binding parameters for ABC graphene used in this work. All numbers are eV, superscript $^a$ denotes parameters that coincide with the ones determined for ABA graphene according to Ref. \onlinecite{zibrov_emergent_2018}.
\label{table1}
}
\end{table}

To determine the parameters to be adjusted, we compare the hopping matrix elements $\gamma_i$ between rhombohedral trilayer graphene and Bernal-stacked trilayer graphene. The dominant parameter $\gamma_0$ corresponds to the hopping between nearest neighbor atoms, specifically the $A$ and $B$ sublattices in a given layer. Parameters $\gamma_1$, $\gamma_3$, and $\gamma_4$ quantify hopping between atoms on adjacent carbon layers. Specifically, for rhombohedral trilayer graphene $\gamma_1$ corresponds to the $B_1$-$A_2$ and $B_2$-$A_3$ hopping matrix element, $\gamma_3$ corresponds to $A_1$-$B_2$ and $A_2$-$B_3$, and $\gamma_4$ corresponds to $A_1$-$A_2$ or $B_2$-$B_3$.  Since Bernal stacked and rhombohedral trilayer graphene are different only in the relative orientation between layers 1 and 3, we expect $\gamma_0$, $\gamma_1$, $\gamma_3$ and $\gamma_4$ parameters to have similar value between these two systems. Therefore, we use the values of Bernal-stacked trilayer graphene in the current calculation, which were matched with experimental results on penetration field capacitance in Ref.~\cite{zibrov_emergent_2018}.

In contrast to the parameters discussed above, $\gamma_2$ in rhombohedral trilayer graphene corresponds to hopping between $A_1$ and $B_3$ atoms, which are located on the outer layers and are on top of each other. In comparison, for Bernal-stacked trilayer those atoms are $A_1$ and $A_3$, so the value of $\gamma_2$ is expected to be different between these two systems. In fact, $\gamma_2$ plays an important role in the current study, since it controls the flatness of conduction band, thus changing the onset of symmetry breaking caused by electron-electron interactions. In what follows we fix $\gamma_2=15\,{\rm meV}$ which gives  the best match of theoretical result based on Stoner model (see below) and experimentally obtained penetration field capacitance, while being close to values reported in previous studies. The remaining two parameters which need to be fixed are $\delta$ and $\Delta_2$.  The convention used for $\delta$ for rhombohedral trilayer graphene is opposite compared to the one used in Bernal-stacked graphene. Namely in Bernal-stacked trilayers $\delta$ is an on-site potential for the sites which are on top of each other and is  positive~\cite{serbyn_new_2013,zibrov_emergent_2018}, while for ABC graphene $\delta$ in Eq.~(\ref{Ham6}) corresponds to an energy shift for atoms $A_1$ and $B_3$. These atoms lack the neighbor in the middle layer and predominantly host the low energy bands, which could explain the negative value of $\delta$ used in the literature~\cite{zhang_band_2010}. As can be seen from the effective two band form of Eq.~\eqref{Ham6} (see Refs.~\cite{koshino_interlayer_2010,zhang_band_2010,jung_accurate_2014,chittari_gate-tunable_2019}), valid for $\mathbf{k}\approx0$, both $\delta$ and $\Delta_2$ act as an energy shift. The value of these two parameters thus mostly modifies the central pocket at low densities, and affects the location of the DOS jump in the valence band. Additionally, $\Delta_2$ modifies the effective mass of particles close to $\mathbf{k}=0$, and changes the value of $\Delta_1$ where a six pocket regime emerges in the valence band. Unfortunately, these are weak effects and cannot be used to fix the value of $\Delta_2$ independent of other parameters. 
Using self-consistent Hartree screening calculation we obtain negative values of $\Delta_2$ of the order of few meV. Based on this considerations, we use experimental data at $\Delta_1=0$ to fix these two parameters and match the location of the ``step'' feature observed at $n_{\rm e}=-1.2\times10^{12}\,\mathrm{cm^2}$ under the constraint that both $\Delta_2, \delta<0$. The resulting values of all parameters used in the current study are listed in Table~\ref{table1}.

\subsection{Compressibility in non-interacting model}
In order to demonstrate the crucial role of interactions, it is instructive to compare the experimental data for inverse compressibility in absence of magnetic field, Fig.~1{\bf d,e}, to the theoretical predictions of the non-interacting model derived from the Hamiltonian of \eqref{Ham6} and the parameters in Table~\ref{table1}.  
We calculate the inverse compressibility $\kappa=\partial \mu/\partial n$ that corresponds to inverse density of states for non-interacting ABC graphene,with the results shown in Fig.~\ref{Fig:NonInteractingDOS}(a) as a function of density of electrons, 
$n_{\rm e}$ and the interlayer potential difference $\Delta_1$ (which is proportional to the displacement field applied in experiment, $D$). 

The features observed in Fig.~\ref{Fig:NonInteractingDOS}a correspond to changes in Fermi surface topology. Figure~\ref{Fig:NonInteractingDOS}(b) show the density of states as a function of carrier density for zero displacement field, $\Delta_1 =0$. 
While density of states has similar features in the electron and hole bands, these features are quantitatively different due to particle-hole symmetry breaking parameters. In particular, at low $\Delta_1$ both conduction and valence band have a van Hove singularity (vHs) in the density of states that arises from the merger of several low energy electron pockets, followed (as the absolute density is increased) by the jump in the density of states due to disappearance of a central pocket. However, in the conduction band the central pocket disappears at densities that are close to the density where vHs occurs, rendering nearly invisible (see Fig.~\ref{Fig:NonInteractingDOS}c-d, which shows Fermi surfaces at several densities). In contrast, in the valence band the DOS jump occurs at much larger hole densities, where it is well resolved.

Large displacement fields, for example $\Delta_1 = 30\,{\rm meV}$ as shown in Figs. \ref{Fig:NonInteractingDOS}e-g, opens a gap at charge neutrality and flattens the bottom of both valence and conduction bands. This manifests in considerably higher values of the DOS at the peaks in  Fig.~\ref{Fig:NonInteractingDOS}e as compared to Fig.~\ref{Fig:NonInteractingDOS}b. Moreover, high displacement field introduces additional features at small hole densities in the valence band:  a Fermi surface with 6 hole pockets emerges with a corresponding jump in the DOS at low density (see middle panel in Fig.~\ref{Fig:NonInteractingDOS}f). In contrast, most conduction band features do not change quantitatively even at strong displacement fields. 

\begin{figure*}[t]
	\begin{center}
		\includegraphics[width=\columnwidth]{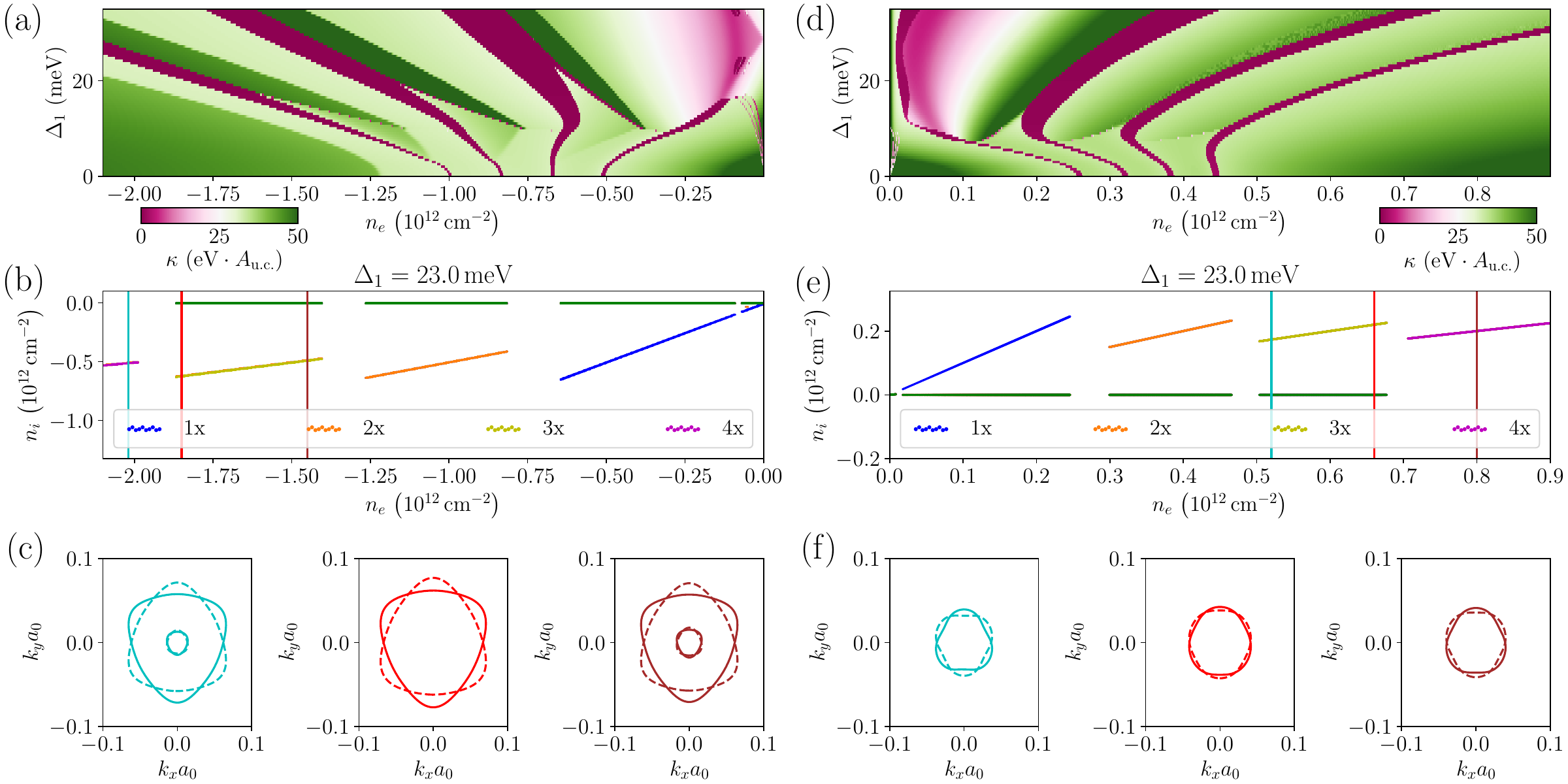}\\
		\caption{
		\label{Fig:InteractingSU4DOS}
			The contour plot of inverse compressibility as a function of density and displacement field $\Delta_1$ with SU(4) symmetric interaction for hole (a) and electron (d) side. Realized phases with possible degeneracy structure (1x,2x,3x and 4x) as a function of density for $\Delta_1=20\,\mathrm{meV}$ (b,e).
			(c,f) show some FS corresponding to densities denoted by vertical lines in (b,e), solid and dashed lines correspond to $K$ and $K'$ valleys. }
	\end{center}
\end{figure*}

\section{Rigid band Stoner model}\label{Sec:Stoner}
\subsection{SU(4) symmetric model}

To gain insight about electron-electron interaction and symmetry broken phases we first turn to a rigid band Stoner model, resembling that used recently for twisted bilayer graphene~\cite{zondiner_cascade_2020}. Taking into account spin and valley degrees of freedom we consider a grand potential per area of the system with four density flavors, namely
\begin{equation}\label{Eq:grand}
	\frac{\Phi_\mathrm{MF}\left(\{\mu_\alpha\}\right)}{A}=\sum_\alpha E(\mu_\alpha)+V_\text{int}-\mu\sum_\alpha n\left(\mu_\alpha\right),
\end{equation}
where $n\left(\mu_\alpha\right)=\int_{0}^{\epsilon}\rho\left(\epsilon\right) d\epsilon$ is the density for a given flavor, $E\left(\mu_\alpha\right)=\int_{0}^{\mu_\alpha}\epsilon\rho\left(\epsilon\right)d\epsilon$ is the total kinetic energy of a given flavor,  and $\rho\left(\epsilon\right)$ is the DOS. 
The term $V_\text{int}$ describes interaction between different flavors, which as a first approximation we take to be SU(4) symmetric:
\begin{equation}
V_\text{int}=\frac{UA_{\rm u.c.}}{2}\sum_{\alpha\neq\beta}n\left(\mu_\alpha\right)n\left(\mu_\beta\right),
\end{equation}
where $U$ is the interaction strength and $A_{\rm u.c.} = (\sqrt{3}/2) a_0^2$ is the area of a unit cell. We fix the total chemical potential and determine the realized phase through minimization of the grand potential with respect to flavor densities. Since for the specific band structure the dependence of $n(\mu)$ is fixed, the variational parameters are either flavor densities $n_\alpha\equiv n\left(\mu_\alpha\right)$ or, equivalently, individual flavor chemical potentials $\mu_\alpha\equiv\mu\left(n_\alpha\right)$. 
Noting that $\partial E\left(\mu_\alpha\right)/\partial n\left(\mu_\alpha\right)=\mu\left(n_\alpha\right)$ the minimization condition for grand potential, Eq.~(\ref{Eq:grand}), reads:
\begin{equation}
\frac{\partial E}{\partial n_\alpha}+UA_{\rm u.c.}\sum_{\beta\neq\alpha}n_\beta-\mu=\mu\left(n_\alpha\right)+UA_{\rm u.c.}\sum_{\beta\neq\alpha}n_\beta-\mu=0.
\label{MinSu4Stoner}
\end{equation}

Rather than solving the above system of four equations for general case, we restrict the search of solution to a symmetric case and a subset of states with broken symmetries. The trivial symmetric solution (which is the ground state for the $U=0$ case) corresponds to the densities of all four flavors being the same. When the Stoner criterion $U\rho\left(\epsilon\right)>1$ is satisfied, we expect the symmetry broken phases to appear. We consider the subset of possible solutions parameterized by a discrete parameter $l$, where $l$ flavors have density $n_{\rm e}$, while remaining $4-l$ flavors have density $n^\prime$ with $n>n^\prime$. Parameter $l$ can take values between 1 and 3, thus labeling a number of filled Fermi surfaces. Corresponding equations determining the densities can be deduced from the general Eq.~\eqref{MinSu4Stoner}. After solving those equations for all values of $l$,  we select the solution with the lowest value of grand potential. Once we have determined the densities for all flavors for specific total chemical potential, we sum all densities together and get the dependence of the density on total chemical potential. Differentiating this function we finally obtain density of states and inverse compressibility.

Figs.~\ref{Fig:InteractingSU4DOS}a and d show a color plot of compressibility, similar to one in Fig.~\ref{Fig:NonInteractingDOS} but for the SU(4) symmetric Stoner model. For both electrons and holes, we observed four distinct regions in which the flavor densities evolve continuously, separated by regions where, in our model, the compressibility is undefined.  Physically, these regions correspond to first order phase transitions where the system breaks up into domains of competing phases, and where negative compressibility is therefore expected. 
For large fillings of valence and conduction band, the system is not polarized and carriers are equally distributed between four flavors. As we move closer to neutrality point the polarization into triple, twofold and single flavor degenerate regions is observed, as can be seen clearly in the  plots of individual densities as a function of a total density for fixed $\Delta_1$ in Figs.~\ref{Fig:InteractingSU4DOS}b and e. The corresponding Fermi surfaces, shown in Fig.~\ref{Fig:InteractingSU4DOS}c and f, reveal the primary difference between conduction and valence band is the Fermi surface topology rather than symmetry breaking.  In the valence band, each flavor polarized region contains a replicas of the DOS jump observed in the single particle band structure. In the conduction band, instead, no such features are observed.  This is expected: interations strongly disfavor occupation of the bands where the density of states diverges, and so such situations are avoided by ferromagnetism polarization.  Because in the conduction band the DOS-step is very close to the van Hove singularity, they are not observed in experiment or in our interacting theory.  

\subsection{Phenomenological breaking of SU(4) symmetry}

In reality, interactions in graphene systems do not have perfect SU(4) symmetry; rather, the true symmetries are separate SU(2) symmetry of the electron spin and a U(1) symmetry relating the two valleys.  Interactions may thus exist which introduce additional scattering channels between the valleys, explicitly breaking the SU(4) symmetry. We account for such symmetry breaking by modifying the interaction potential in the rigid Stoner model as
\begin{equation}
V_\text{int}=\frac{UA_{\rm u.c.}}{2}\sum_{\alpha\neq\beta}n_\alpha n_\beta+JA_{\rm u.c.}(n_1-n_3)(n_2-n_4),
\label{IntSU4Break}
\end{equation} 
where  we take the following correspondence between values of Greek indices and valley and spin orientations: $1=\{K,\uparrow\}$, $2=\{K^\prime,\uparrow\}$, $3=\{K,\downarrow\}$, $4=\{K^\prime,\downarrow\}$, and $J<0$ so that the lowest energy two-fold states are spin polarized and valley unpolarized as observed in experiment. The second term that is proportional to $J$ physically corresponds to a Hund's rule type anisotropy, that will be justified on a microscopic basis in the next section. 

The effect of the $J$ term is twofold. As is clear from Eq.~(\ref{IntSU4Break}) the four flavors split into non-equivalent groups of two. Therefore, increasing $|J|$ leads to suppression of three-fold degenerate and one fold degenerate regions in the phase diagram. In addition, $|J|>0$ allows for existence of more complicated solutions such as $n_1=0$, $n_3=n_4>n_2>0$. Since the Stoner model calculation is performed on realistic band structure of rhombohedral trilayer graphene, the realized phases are highly dependent of band structure parameters.  

\section{Hartree-Fock model}
To test the robustness of the results obtained from the simple Stoner
model, we perform a more detailed Hatree-Fock analysis of the cascade
of spin and valley symmetry-breaking transitions as a function of
density. We begin with the Hamiltonian $H_{0}$ in Eq. \eqref{Ham6}, and add
to it a screened Coulomb interaction
\begin{equation}
H_{{\rm c}}=\frac{1}{2}\sum_{\alpha,\beta,i,j}\int d^{2}rd^{2}r'U(|\bm{r}-\bm{r}'|)\psi_{i\alpha}^{\dagger}(\bm{r})\psi_{j\beta}^{\dagger}(\bm{r}')\psi_{\beta j}(\bm{r}')\psi_{i\alpha}(\bm{r}),\label{eq:H_c}
\end{equation}
where $\alpha,\beta=1,\dots,4$ runs over the spin and valley indices,
$i,j=1,\dots,6$ label the two sublattices and three layers (this
is the six-dimensional space in which $H_{0}$ is written),
and $\psi_{j\alpha}^{\dagger}(\bm{r})$ creates an electron at position
$\bm{r}$, sublattice/layer $j$, and spin/valley $\alpha$. The interaction
is taken to be $U(r)=\int\frac{d^{2}q}{(2\pi)^{2}}U(q)e^{i\bm{q}\cdot\bm{r}}$,
where 
\begin{equation}
{U}(q)=\frac{2\pi e^{2}\tanh qd}{\epsilon q}, \label{eq:Uq}
\end{equation}
corresponding
to screening by two symmetric metallic gates at a distance $d$ above
and below the system ($\epsilon$ is an effective dielectric constant,
which includes the HBN and the screening from the remote bands of the
trilayer graphene). 

The screened Coulomb potential is assumed to depend only on the local
density of electrons, and does not depend on any of the internal indices
(such as spin and valley). This type of interaction is expected to
dominate when the average distance between charge carriers is much
larger than the inter-atomic spacing, as is always the case in our
system. Later, we will include phenomenologically the effects of short-distance
interactions that may depend on the spin and valley indices. 

We treat the Hamiltonian $H=H_{0}+H_{{\rm c}}$ within the Hartree-Fock
approximation. To this end, we use a variational quadratic Hamiltonian
$H_{{\rm var}}$ to generate a variational state, and minimize the
variational free energy over the parameters of $H_{{\rm var}}$. In
order to simply the calculation, we restrict the variational state
in several ways: 1. All the bands away from the Fermi level are ``frozen'',
i.e., we do not allow the variational Hamiltonian to mix the active
(partially filled) band with the completely filled or completely empty
bands; 2. We allow only states that preserve the separate spin and
valley conservation symmetries of the Hamiltonian. The first simplification
is justified as long as the remote bands are separated from the active
band by a sufficiently large gap. Most crucially, it requires the
displacement field to be sufficiently large compared to the magnitude
of the Coulomb interaction. The second simplification omits the possibility
of states that break the valley conservation symmetry spontaneously
by occupying states that are coherent superpositions of the two valleys.
We leave the study of such states to future theoretical work. 

\begin{figure}[t]
\centering
\includegraphics[width=0.45\textwidth]{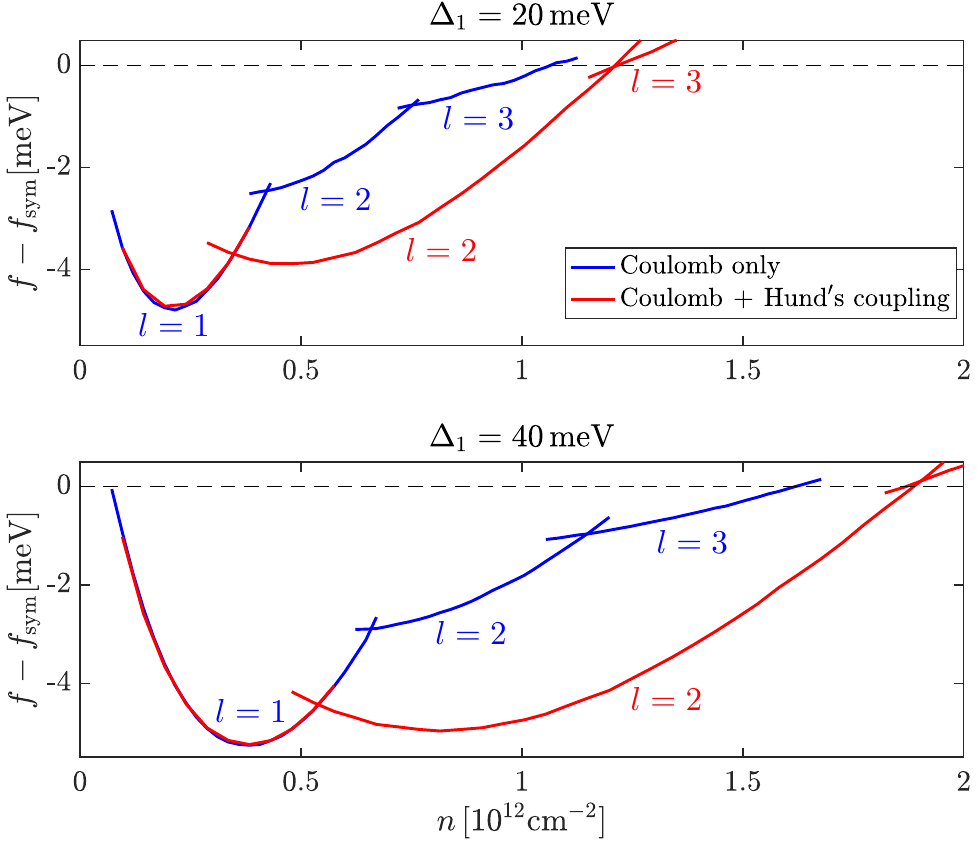}
\caption{Free energy $f$ per electron as a function of electron density, measured relative to that of a spin and valley-symmetric state, $f_{\rm{sym}}$. The blue curve shows the result of a calculation with only a long-range, spin and valley-symmetric screened Coulomb interaction (Eq.~\eqref{eq:H_c}). The red curve was obtained by including also an interaction of the form (\ref{eq:HA}) with $\tilde{J} = -0.03{U}(q=0)$ (see Eq.~\eqref{eq:Uq}), $\tilde{U}=\tilde{V}=0$. The top (bottom) panel shows results for a displacement field of $\Delta_1=20$meV ($\Delta_1=40$meV), respectively. The label $l$ indicates the number of populated spin and valley flavors. The effective dielectric constant was taken to be $\epsilon=8$, and the distance to the metallic gates is $d=150a_0$, where $a_0$ is the unit cell size of graphene. The calculations were performed at a low but finite temperature, $T=0.2$\,meV.}
\label{fig:S:HF}
\end{figure}

The blue curves in Fig.~\ref{fig:S:HF} show representative results for the variational Hatree-Fock free energy per electron
as a function of the electronic density of different types of spin--
and valley--polarized states, measured relative to the symmetric
state (where all the spin and valley flavors are populated equally).
As in the Stoner model, the states labelled by $l=1,2,3$ are states
in which $l$ spin or valley flavors are populated, and the
other $4-l$ flavors are empty. The results are qualitatively
similar to those of the Stoner model, in that the system undergoes
a cascade of first-order transitions upon increasing the density from
the charge neutrality point, according to the sequence (1,2,3,4). Qualitatively similar results are also obtained for other values of the displacement field $\Delta_1$ and for a hole-doped system. 

In order to match the experiment, where the $l=3$ state is not observed,
we need to consider interaction terms beyond the long-range Coulomb
interaction [Eq. \eqref{eq:H_c}] which is SU(4) symmetric in spin and valley
space. Interactions at ranges of the order of a few lattice constants
(either Coulombic or phonon-mediated) can depend on the spin and valley
indices of the electrons. We model these short-range
terms as
\begin{equation}
H_{{\rm A}}=\frac{\tilde{U}}{2}\int d^{2}r:\!(n_{K}+n_{K'})^{2}\!:+\frac{\tilde{V}}{2}\int d^{2}r:\!(n_{K}-n_{K'})^{2}\!:+\tilde{J}\int d^{2}r:\!\bm{\sigma}_{K}\cdot\bm{\sigma}_{K'}\!:,
\label{eq:HA}
\end{equation}
where $n_{K}(\bm{r})$ and $n_{K'}(\bm{r})$ are the local densities
of electrons in the two valleys.  The operators $\bm{\sigma}_{K}(\bm{r})$, $\bm{\sigma}_{K'}(\bm{r})$
are the local spin densities, defined as $\bm{\sigma}_{K}(\bm{r})=\sum_{s_{1},s_{2}}\psi_{Ks_{1}}^{\dagger}(\bm{r})\bm{\sigma}_{s_{1},s_{2}}\psi_{Ks_{2}}(\bm{r})$ (and similarly for $\bm{\sigma}_{K'}$),
where $\bm{\sigma}_{s,s'}$ are Pauli matrices acting in spin space, and
$\psi_{Ks}^{\dagger}$ is the creation operator for electrons at valley
$K$ and spin $s=\uparrow,\downarrow$ in the active band that crosses
the Fermi level (which we project to in our calculation). Normal ordering of an operator $\hat{O}$ is denoted by $:\!\hat{O}\!:$.
The $\tilde{U}$ term is a local
spin and valley-isotropic interaction, $\tilde{V}$ parameterizes the difference in the local interaction strength between electrons of the same valley and electrons in different valleys, and the $\tilde{J}$ term is a spin-exchange Hund's rule coupling between electrons of different valleys. Eq.~\eqref{eq:HA} is the most general contact (zero-range) interaction term between electrons in the active band, which is compatible with all the symmetries of the problem. 

Our parameterization of the short-range interactions is related to that used
by Kharitonov~\cite{kharitonov_phase_2012}  in modeling the spin and valley polarization
of electrons in the zero Landau levels in graphene,
\begin{equation}
H_{{\rm A}}=\int d^{2}r\left[\frac{u_{z}}{2}:\!(\psi^{\dagger}\tau^{z}\psi)(\psi^{\dagger}\tau^{z}\psi)\!:+u_{\perp}:\!(\psi^{\dagger}\tau^{+}\psi)(\psi^{\dagger}\tau^{-}\psi)\!:+H.c.\right],
\end{equation}
where $\tau$ are Pauli matrices that act in valley space and $\psi^{\dagger}=(\psi_{K\uparrow}^{\dagger},\psi_{K\downarrow}^{\dagger},\psi_{K'\uparrow}^{\dagger},\psi_{K'\downarrow}^{\dagger})$,
by the following relations: $\tilde{U}=-\frac{u_{\perp}}{4}$, $\tilde{V}=u_{z}-\frac{u_{\perp}}{4}$,
$\tilde{J}=-\frac{u_{\perp}}{2}$. 

We have explored the Hartree-Fock phase diagram of the system including the short-range interaction term (\ref{eq:HA}) for a few parameter sets, leaving a complete mapping of the phase diagram for future work. The red curves in Fig.~\ref{fig:S:HF} show the free energy as a function of electronic density with an additional ferromagnetic Hund's rule coupling of strength $\tilde{J}=-0.03\, {U}(q=0)$ (with ${U}(q)$ given by Eq.~\eqref{eq:Uq}). The other parameters in Eq.~\ref{eq:HA}, $\tilde{U}$ and $\tilde{V}$, were set to zero. As can be seen in the figure, this value of the Hund's coupling is sufficient to completely suppress the $l=3$ phase, in agreement with the experiment. Since $\tilde{J}<0$, the region labelled by $l=2$ is a valley-unpolarized, spin-polarized phase; this phase is favored by the $J$ term compared to the $l=3$ region, causing the disappearance of the latter phase. For $|\tilde{J}|<0.03 {U}(0)$, the $l=3$ region reappears. The $l=2$ region has the same spin polarization per electron as the $l=1$ region, in agreement with experiment (see Figs. \ref{fig2} and \ref{fig3} of the main text). 

The $l=3$ region can also be suppressed by increasing $\tilde{V}$ while keeping $\tilde{U}=-\tilde{V}$, which increases the repulsion between electrons in the same valley relative to the repulsion between electrons of opposite valleys. However, even for $\tilde{V}$ as strong as $0.25\, U(q=0)$, the $l=3$ region is not completely eliminated. Thus, within our model, we conclude that the inter-valley Hund's rule coupling $\tilde{J}$ is much more effective than the $\tilde{V}$ term in suppressing the $l=3$ region.

The anisotropic interaction terms that are necessary to suppress the 3-fold region in our calculations originate from lattice scale Coulomb repulsion or electron-phonon coupling. The natural scale for such terms can be estimated as $\int^{a_0}_0 d^2r e^2 / (\epsilon r) = 2\pi e^2 a_0/\epsilon$, where $a_0 = 0.246$\,nm is the graphene unit cell size and $\epsilon$ is an effective dielectric constant, taken for simplicity to be identical to that used in the long-range Coulomb interaction (\ref{eq:Uq}). Since we have used $d = 150 a_0 \approx 40$\,nm in our calculations, the short-range interactions are naively estimated to be of the order of $0.5\%-1\%$ of ${U}(q=0)$. The Hund's rule term needed to suppress the $l=3$ region is of this order of magnitude (although somewhat larger).

It is interesting to note that the current experiment strongly suggests that the 2-fold region is fully spin polarized, indicating that $\tilde{J}<0$. In contrast, in the $\nu=0$ quantum Hall insulator in both monolayer and bilayer graphene, experiments indicate that the ground state is a valley-unpolarized, nearly spin-unpolarized state (a ``canted anti-ferromagnet''~\cite{maher_evidence_2013,young_tunable_2014}), corresponding to $\tilde{J}>0$. The difference between the two may come from the strong renormalization of the short-range anisotropic interactions from remote bands, which may be very different in our system than in monolayer and bilayer graphene in the quantum Hall regime.

\end{document}